\documentclass[a4paper,11pt]{article}
\usepackage{jheppub} 

\usepackage{amsmath}
\usepackage{bbm}
\usepackage{comment}
\usepackage{graphicx,enumerate,xcolor,enumitem,xspace,braket,mathrsfs} 
\usepackage{soul}
\allowdisplaybreaks[1]

\newcommand{\tr}{\mathrm{Tr}}

\usepackage[normalem]{ulem}

\begin{document}
\preprint{ PITT-PACC-2604}
\title{Spin Correlation and Quantum Entanglement of Fermion Pairs in Transversely Polarized $e^-e^+$ Collisions}
\author[a,b]{Yi-Jing Fang,}
\author[c]{Amit Bhoonah,}
\author[c]{Kun Cheng,}
\author[c]{Tao Han,}
\author[d,e]{Yandong Liu,}
\author[a,b,f]{and Hao Zhang}

\affiliation[a]{Theoretical Physics Division, Institute of High Energy Physics, Chinese Academy of Sciences, Beijing 100049, China}
\affiliation[b]{School of Physics, University of Chinese Academy of Sciences, Beijing 10049, China}
\affiliation[c]{Pitt PACC, Department of Physics and Astronomy,\\ University of Pittsburgh, 3941 O’Hara St., Pittsburgh, PA 15260, USA}
\affiliation[d]{School of Physics and Astronomy, Beijing Normal University, Beijing, 100875, China}
\affiliation[e]{Key Laboratory of Multiscale Spin Physics (Ministry of Education), Beijing Normal University, Beijing, 100875, China}
\affiliation[f]{Center for High Energy Physics, Peking University, Beijing 100871, China}

\emailAdd{fangyj@ihep.ac.cn}
\emailAdd{aab266@pitt.edu}
\emailAdd{kun.cheng@pitt.edu}
\emailAdd{than@pitt.edu}
\emailAdd{ydliu@bnu.edu.cn}
\emailAdd{zhanghao@ihep.ac.cn}

\date{\today}

\abstract{
We systematically study the spin correlations and quantum entanglement in transversely polarized electron-positron collisions.  
We find that the $s$-channel QED process $e^-e^+\to f\bar f$ produces a maximally entangled state in the entire phase space when the initial beams are transversely polarized, while the quantum magic varies in different phase space points for the maximally entangled Bell states. 
For electroweak processes, the spin configuration of final states depends on chiral couplings, and the entanglement is also greatly enhanced by transverse polarization as in the QED process.
For Bhabha scattering with additional $t$-channel contributions,  the transverse polarization still increases the final state entanglement, although with some dilution.
The sensitive dependence of final spin states on the transverse polarization makes the beam polarization a powerful tool for generating and controlling quantum entanglement in collider experiments, opening up new opportunities for quantum information studies at high-energy colliders. 
}

\maketitle

\section{Introduction}

Quantum mechanics forms the basis of modern physics, and entanglement~\cite{Schrödinger_1935,PhysicsPhysiqueFizika.1.195,PhysRev.47.777} stands out as one of its most distinct phenomena. In recent years, there has been increasing interest in studying quantum correlations and entanglement in high-energy collider experiments~\cite{Afik:2020onf,Barr:2024djo}. For example, a pair of spin-1/2 particles produced at colliders can be treated as a two-qubit quantum system, enabling the study of quantum information. 
Among the various systems investigated at colliders, the top quark pair has been the most extensively studied~\cite{Afik:2020onf,Fabbrichesi:2021npl,Severi:2021cnj,Afik:2022kwm,Afik:2022dgh,Han:2023fci,Aguilar_Saavedra_2022,Dong_2024,Cheng:2023qmz,Cheng:2024btk,Han:2024ugl,Afik:2026pxv}, and quantum entanglement has been observed at the highest energy with the top pair produced at the Large Hadron Collider~\cite{ATLAS:2023fsd,CMSCollaboration_2024}. Given the large amount of quantum data produced in collider experiments, there is a plethora of processes and systems that can be constructed to study entanglement and other quantum information. Examples include 
$\tau$ lepton pairs \cite{Altakach:2022ywa,Ehataht:2023zzt,Ma:2023yvd,Fabbrichesi:2024wcd,Han:2025ewp,Zhang:2025mmm,Ai:2025wnt}, massive gauge boson pairs~\cite{Barr:2021zcp,Barr:2022wyq,Ashby-Pickering:2022umy,Aguilar-Saavedra:2022wam,Fabbrichesi:2023cev,Fabbri:2023ncz,Bi:2023uop,Bernal:2024xhm,Grossi:2024jae,Goncalves:2025qem,Goncalves:2025xer}, $\Lambda$ baryons~\cite{Du:2024sly,Afik:2025grr,Chen:2013epa,Fabbrichesi:2024rec,Pei:2025yvr,Pei:2025ito,Fucilla:2025kit} and light quark pairs~\cite{Cheng:2025cuv,Lin:2025eci,Cao:2025qua}.
The quantum information of these particle pairs produced via various new physics effects has also been investigated~\cite{Aoude:2022imd,Fabbrichesi:2022ovb,Severi:2022qjy,Fabbrichesi:2023jep,Bernal:2023ruk,Aoude:2023hxv,Maltoni:2024tul,Altakach:2022ywa,Du:2024sly,Fabbrichesi:2025ywl,Aoude:2025jzc,Cao:2025qua}, and the impact of scattering kinematics on final quantum states has been extensively analyzed~\cite{Cheng:2024btk,Cheng:2024rxi}. 
Quantum tomography in quark flavor transitions for neutral meson and anti-meson systems has also been formulated \cite{Cheng:2025zcf}. 
Overall, the quantum states produced at colliders exhibit diverse configurations depending on the initial conditions, interactions, and scattering kinematics. The broad variety of available systems makes high-energy colliders a rich laboratory for exploring quantum information science.

Given the clean experimental environment in $e^-e^+$ collisions and the excellent control of the colliding energy and luminosity, as well as the possible high degree of beam polarizations, an $e^-e^+$ collider may be an ideal playground for constructing quantum information in a fully controlled manner. 
In this work, we utilize the beam polarization of the initial state $e^\pm$. Although the effect of longitudinal polarization has recently been studied \cite{Altakach:2026fpl,Guo:2026yhz},  we demonstrate that transverse polarizations of initial states have more unique and dramatic effects.\footnote{During the completion of this work, a manuscript 
\cite{Zhang:2026nwm} appeared showing certain transverse polarization effects for hyperon production and decay at low-energy QED.}

It is important to point out that,  if both initial beams are purely transversely polarized in the $s$-channel QED process $e^-e^+\to f\bar{f}$, we find that the spin state of the final fermion pair $\ket{f\bar f}$ is a pure Bell state, maximally entangled \textit{at any scattering angle and scattering energy}.
This stands in stark contrast to the commonly known result that, with unpolarized or longitudinally polarized beams, large entanglement only occurs when the final states are both highly boosted and centrally scattered.
The dramatic effect of the transverse polarization arises primarily because it allows the interference between amplitudes with different initial state helicities~\cite{Wen:2023xxc,Cheng:2025zaw}. A transversely polarized state is a superposition of left- and right-handed polarizations along the longitudinal direction. For example, the spin state of  $e^-e^+$ beams transversely polarized along the $x$-axis is given by
\begin{align}\label{eq:statevector_intro}
    &\frac{1}{\sqrt{2}}(\ket{R}+\ket{L})\otimes
    \frac{1}{\sqrt{2}}(\ket{R}+\ket{L})  = \frac{1}{2}\big(\ket{RL}+\ket{RR}+\ket{LL}+\ket{LR}\big), 
\end{align}
where $R/L$ denotes the right/left helicity of the incoming $e^\pm$.
In the massless limit, only the $e^+$ and $e^-$ with opposite helicities (parallel spins) annihilate each other for gauge interactions. Consequently, the coherent superposition $\ket{RL}+\ket{LR}$ in Eq.~\eqref{eq:statevector_intro}, a maximally entangled part of the initial state, is selected by the vector current interaction.  The fermion pair produced from $e^-e^+$ annihilation is therefore guaranteed to be maximally entangled if the left- and right-handed coupling strengths are equal.
For electroweak $s$-channel processes involving $Z$-boson exchange, such as $e^-e^+\to t\bar{t}/\tau^-\tau^+$, the entanglement behavior of the final states usually resembles that of the QED process, up to minor corrections from chiral couplings, scaled by the ratio of the axial-current and vector-current couplings $g_A/g_V$.

The unique final quantum states exhibit certain features of quantum information.  As a representative example, we present the quantum entanglement and quantum magic of the state for a comparative study. Entanglement characterizes the non-classical correlation of a state that arises from a quantum superposition. Magic, or non-stabilizerness, is the departure of a state from the set of stabilizer states generated by Clifford gates; these stabilizer states can be efficiently simulated on classical computers~\cite{Nielsen:2012yss}. 

The transverse polarization of beams naturally exists in a circular accelerator due to Sokolov–Ternov effects~\cite{Sokolov:1963zn}, and the transverse polarization would also be an intermediate stage for preparing longitudinally polarized beams~\cite{CEPCStudyGroup:2023quu,Accardi:2012qut}. In realistic situations where the initial states are not 100\% transversely polarized, the final state is a mixed state that is partly composed of a Bell state, the percentage of which is determined by the degree of polarization. The dramatic dependence of quantum entanglement on the polarization of initial beams also provides the possibility of manipulating the quantum state produced at colliders.

The remainder of the paper is organized as follows. Section~\ref{Section:General framework} reviews the general features of two-qubit systems and the quantum states produced at colliders. In Section~\ref{section:toymodel}, we start from the QED process to explore the spin state and entanglement of the final state produced from transversely polarized beams, obtaining our main result that the annihilation of transversely polarized leptons always produces Bell states regardless of the scattering kinematics. 
We generalize our discussion in Section~\ref{section:Spin_correlation} to electro-weak processes with chiral interactions, and show that the final state depends crucially on the relation between vector and axial couplings.
Finally, we summarize in Section~\ref{section:summary} and discuss some extensions with regard to the $t$-channel process and fictitious states.

\section{General Framework}
\label{Section:General framework}

\subsection{Spin density matrix and quantum entanglement}

In quantum mechanics, a pure state can be described by a state vector $\ket{\psi}$. 
Generally, when a quantum state is an incoherent mixture of different pure states $\ket{\psi_i}$, this mixed state should be described by a density operator,
\begin{equation}
    \hat\rho_{\rm{mixed}}=\sum_{i=1}^np_i\ket{\psi_i}\bra{\psi_i}, ~~p_i>0, ~~\sum_{i=1}^np_i=1,
    \label{eq:mixed}
\end{equation}
where $p_i$ is the probability fraction of the pure state $\ket{\psi_i}$.

Quantum entanglement (non-separability) is a criterion to characterize the non-classical correlations predicted by quantum mechanics.
Consider a bipartite system composed of two qubits, $a$ and $b$, in the Hilbert space $\mathcal{H}=\mathcal{H}_a\otimes\mathcal{H}_b$.
A pure state is separable if the state vector can be written as $\ket{\psi}=\ket{\psi^a}\otimes\ket{\psi^b}$, otherwise the state is entangled.
For general mixed states, a separable quantum state can be written as a convex combination of product states, which can be mathematically expressed as~\cite{PhysRevLett.77.1413,HORODECKI1997333}: 
\begin{equation}\label{eq:separable}
    \hat\rho=\sum_np_n{\hat\rho}_n^a\otimes{\hat\rho}^b_n, \quad \sum_np_n=1, \quad p_n > 0,
\end{equation}
where $\hat\rho^{a,b}_n$ are density operators in the sub-systems $a$ and $b$. On the other hand, a quantum state that \emph{cannot} be written in the direct product form of Eq.~\eqref{eq:separable} is said to be \emph{non-separable} and thus entangled.

One quantitative measure of entanglement is the concurrence~\cite{PhysRevLett.80.2245}. For a two-qubit system, it is defined as
\begin{equation}
    \mathscr{C}\left[{\rho}\right]\equiv \max\left(0, \lambda_1-\lambda_2-\lambda_3-\lambda_4\right), \qquad 0\ {\leqslant}\ \mathscr{C} [{\rho}]\ {\leqslant} \ 1,
    \label{eq:Concurrence}
\end{equation}
where $\lambda_i$ are the eigenvalues of the operator $\sqrt{\sqrt{{\rho}}\tilde{{\rho}}\sqrt{{\rho}}}$ ordered in decreasing magnitude, $\tilde{{\rho}}=\left(\sigma_2\otimes\sigma_2\right){\rho}^*\left(\sigma_2\otimes\sigma_2\right)$, and ${\rho}^*$ is the complex conjugate of $\rho$. A state is entangled \emph{if and only if} $\mathscr{C}[{\rho}]>0$, with $\mathscr{C}[{\rho}] = 1$ indicating that the state ${\rho}$ is a Bell state with maximum entanglement.

For an intuitive understanding of how two spin-1/2 particles are polarized and correlated along different spatial directions, it is often convenient to decompose the $4\times4$ density matrix of the density operator in a given basis using Pauli matrices,
\begin{equation}\label{eq:rhoCij}   \rho=\frac{\mathbb{I}_2\otimes\mathbb{I}_2+B_i^a\sigma^i\otimes\mathbb{I}_2+B_i^b\mathbb{I}_2\otimes\sigma^i+C_{ij}\sigma^i\otimes\sigma^j}{4},
\end{equation}
where the sum over $i,j=1,2,3$ is implicit, and $\mathbb{I}_2$ is the $2\times2$ identity matrix. Here, $B_i^{a}$ and $B_j^{b}$ are the net polarizations of particles $a$ and $b$, respectively, while $C_{ij}$ is the spin correlation between the two particles.  As such, the quantum state is fully specified by these 15 parameters.
For example, there exist four linearly independent Bell states in a two-qubit system. In an orthonormal basis $\{\hat{e}_1,~\hat{e}_2,~\hat{e}_3\}$, their state vectors and spin correlation matrices are represented as
\begin{subequations}
 \label{eq:Bell_state}
\begin{align}
    &\ket{\Psi_0}=\frac{\ket{\uparrow\downarrow}-\ket{\downarrow\uparrow}}{\sqrt{2}},\quad B_i^{a/b}=0, \quad ~C_{ij}=\begin{pmatrix}
        -1 &0 &0 \\
        0 &-1 &0 \\
        0 &0 &-1 \\
    \end{pmatrix}_{ij},\\
    &\ket{\Psi_1}=\frac{\ket{\uparrow_{\hat{e}_1}\downarrow_{\hat{e}_1}}+\ket{\downarrow_{\hat{e}_1}\uparrow_{\hat{e}_1}}}{\sqrt{2}},\quad B_i^{a/b}=0, \quad ~C_{ij}=\begin{pmatrix}
        -1 &0 &0 \\
        0 &1 &0 \\
        0 &0 &1 \\
    \end{pmatrix}_{ij},\\
    &\ket{\Psi_2}=\frac{\ket{\uparrow_{\hat{e}_2}\downarrow_{\hat{e}_2}}+\ket{\downarrow_{\hat{e}_2}\uparrow_{\hat{e}_2}}}{\sqrt{2}},\quad B_i^{a/b}=0, \quad ~C_{ij}=\begin{pmatrix}
        1 &0 &0 \\
        0 &-1 &0 \\
        0 &0 &1 \\
    \end{pmatrix}_{ij},\\
    &\ket{\Psi_3}=\frac{\ket{\uparrow_{\hat{e}_3}\downarrow_{\hat{e}_3}}+\ket{\downarrow_{\hat{e}_3}\uparrow_{\hat{e}_3}}}{\sqrt{2}},\quad B_i^{a/b}=0, \quad ~C_{ij}=\begin{pmatrix}
        1 &0 &0 \\
        0 &1 &0 \\
        0 &0 &-1 \\
    \end{pmatrix}_{ij},
\end{align} 
\end{subequations}
where $\ket{\uparrow_{\hat{e}_i}}(\ket{\downarrow_{\hat{e}_i}})$ denotes the spin-up(down) eigenstate quantized along the $\hat{e}_i$ direction ({$i,j=1,2,3$). 
Here, $\ket{\Psi_0}$ is a spin singlet state $\ket{J=\mathbf{0},J_z=0}$ and has the same form under any choice of basis. The others are the triplet states $\ket{J=\mathbf{1}, J_z=0}$, with $J_z$ along the three different axes $\hat e_i$.
More generally with respect to an arbitrary spatial direction $\hat{a}$, we use $\ket{\Psi}_{\hat a}$ to denote the  triplet state $\ket{J=\mathbf{1},(\vec J\cdot\hat a)=0}$ in the rest of this paper. Explicitly, its state vector and spin correlation matrix are
\begin{equation}
 \ket{\Psi}_{\hat a}\equiv\frac{\ket{\uparrow_{\hat a}\downarrow_{\hat a}}+\ket{\downarrow_{\hat a}\uparrow_{\hat a}}}{\sqrt{2}},\qquad C_{ij}=\delta_{ij}-2a_i a_j.
\end{equation}

The representation in Eq.~\eqref{eq:rhoCij} also makes the entanglement more transparent and provides a simpler form of concurrence when the polarization vector $B_i^{a/b}=0$.  In this case, the concurrence is determined from the three eigenvalues $c_i$ of $C_{ij}$:
\begin{equation}
    \mathscr{C}[{\rho}]=\max\left(0,\frac{-1-\tr(C)}{2}, \frac{\tr(C)-2c_{i}-1}{2} \right).
\end{equation}
The maximum entanglement is reached for the concurrence $\mathscr{C}[{\rho}]=1$ only when all eigenvalues maximize with $|c_i|=1$.
From Eq.~\eqref{eq:Bell_state}, we also see that an incoherent mixture among different Bell states greatly reduces the concurrence. This could happen for unpolarized scattering as will be discussed in Sections~\ref{subsec:tautau} and \ref{subsec:bb}.

\subsection{Quantum states at lepton colliders}
Consider a scattering process 
\begin{equation}
e^-_{\lambda} \ e^+_{\bar{\lambda}}\to f_{\alpha}(k_f) \ \bar{f}_{\bar\alpha}(k_{\bar{f}}),
\end{equation}
where $f\bar{f}$ is the final state fermion pair such as $t\bar{t}$ and $\tau^-\tau^+$.  The spin indices $\lambda/\bar{\lambda}$ denote the polarizations of $e^-/e^+$, and $\alpha/\bar{\alpha}$ that of $f/\bar{f}$. In the center-of-mass (c.m.) frame of $f\bar{f}$, we denote the momenta of $f$ and $\bar{f}$ as
\begin{equation}
    k_f^\mu=(E,\mathbf{k}), \quad k_{\bar{f}}^\mu=(E,-\mathbf{k}).
\end{equation}
If the initial state $\ket{e^-_{\lambda}e^+_{\bar{\lambda}}}$ is a pure state, the final state is also a pure state.\footnote{In this work, we ignore the effects of the initial state and final state radiation \cite{Aoude:2025ovu,Gu:2025ijz}.} Its state vector $\ket{f\bar{f}}$ can be expanded by the eigenstates of momentum and spin $\ket{\mathbf{k},\alpha\bar{\alpha}}\equiv\ket{\mathbf{k},\alpha}\otimes \ket{-\mathbf{k},\bar \alpha}$, 
\begin{align}
   \ket{f\bar{f}} \propto &~\hat{\mathcal{T}}\ket{e^-_\lambda e^+_{\bar\lambda}}
   \nonumber\\
   = & \int d^3\mathbf{k}\sum_{\alpha,\bar \alpha}\ket{\mathbf{k},\alpha\bar{\alpha}}\bra{\mathbf{k},\alpha\bar{\alpha}}\hat{\mathcal{T}}\ket{e^-_\lambda e^+_{\bar\lambda}}=\int d^3\mathbf{k}\sum_{\alpha,\bar \alpha}\ket{\mathbf{k},\alpha\bar{\alpha}}\mathcal{M}_{\alpha\bar\alpha}^{\lambda\bar\lambda}(\mathbf{k}),
   \label{eq:finial_state}
\end{align}
where $\hat{\mathcal{T}}$ is the transition operator and $\mathcal{M}_{\alpha\bar\alpha}^{\lambda\bar\lambda}{(\mathbf{k})}$ is the scattering amplitude.

If the initial state is a mixed state described by 
$\hat\rho^{e^-}\otimes\hat\rho^{e^+}$,\footnote{It is taken into account that the $e^+$ beam and the $e^-$ beam are prepared separately with no spin correlation at all.} generally the final spin state is also a mixed state given by $\hat{\mathcal{T}}(\hat\rho^{e^-}\otimes\hat\rho^{e^+})\hat{\mathcal{T}}^\dagger$; see Ref.~\cite{Cheng:2024btk} for details. 
At a given scattering angle, the (unnormalized) spin density operator is
\begin{equation}
    {{\hat R}}_{f\bar{f}}(\mathbf{k})= 
    \mathcal{M}_{\alpha\bar\alpha}^{\lambda\bar\lambda}(\mathbf{k})\ \rho^{e^-}_{\lambda\lambda^\prime}\ \rho^{e^+}_{\bar\lambda\bar\lambda^\prime}\ (\mathcal{M}_{\alpha'\bar\alpha'}^{\lambda'\bar\lambda'}(\mathbf{k}))^{{*}}\ \ket{\alpha\bar{\alpha}}\bra{\alpha'\bar{\alpha}'}.
    \label{eq:mixed_state}
\end{equation}
 Here, $\rho^{e^-}_{\lambda\lambda^\prime}$ ($\rho^{e^+}_{\bar\lambda\bar\lambda^\prime}$) is the spin density matrix of the initial electron (positron) beam, and the repeated spin indices are summed over.
 The normalized spin density operator of $f\bar f$ is 
 \begin{equation}
  \hat\rho_{f\bar f}(\mathbf{k})=\frac{\hat R_{f\bar f}(\mathbf{k})}{\tr(\hat R_{f\bar f}(\mathbf{k}))},
 \end{equation}
with $\tr({\hat R}_{f\bar f}(\mathbf{k}))$ proportional to the differential cross section.
We can see that the spin density operator of the final state is determined by both the initial spin states and the helicity amplitudes.

\subsection{Notation and basis choice}

The explicit expressions of the amplitudes $\mathcal{M}^{\lambda\bar\lambda}_{\alpha\bar\alpha}$ depend on the choice of the spin quantization axis.
 Throughout this work, we use the helicity $R/L$ to label the spin indices $\lambda/\bar\lambda$ of the initial $e^\pm$ beams.  Only $e^- e^+$ with opposite helicities (parallel spins) interact with each other, and the spin index  $\lambda\bar\lambda=RL$/$LR$ corresponds to the right-/left-handed interaction of the initial states.

For the spin indices $\alpha\bar\alpha$ of the final state $f\bar f$, we choose the same spin quantization axis for both $f$ and $\bar f$ as a combined $f\bar f$ system, rather than their individual helicities, so that it is easier to match it to the spin singlet/triplet states in Eq.~\eqref{eq:Bell_state}.
There are several conventional basis choices for the spin quantization axes. A widely used option is the helicity basis, defined in the c.m. frame as $\{\hat{r},\hat{n},\hat{k}\}$; see Fig.~\ref{fig:basis}(a), where $\hat{k}$ is the momentum direction of fermion $f$, $\hat{r}$ is the vector orthogonal to $\hat{k}$ within the scattering plane defined as
$\hat{r}=-(\hat{z}-\hat{k}\cos{\theta})/\sin{\theta}$ with $\hat{z}$ being the momentum direction of initial beam $e^-$ and  $\theta$ being the scattering angle, and $\hat{n}$ is perpendicular to the scattering plane defined as $\hat{n}=\hat{k}\times\hat{r}$. 
Spin bases are quantized along the 
$\hat k$ direction as 
$\ket{\uparrow_k}$ and 
$\ket{\downarrow_k}$, and their relative phases are determined by the $\hat{r}$ and $\hat{n}$ directions.
\begin{figure}
	\centering     
    \includegraphics[width=0.4\linewidth]{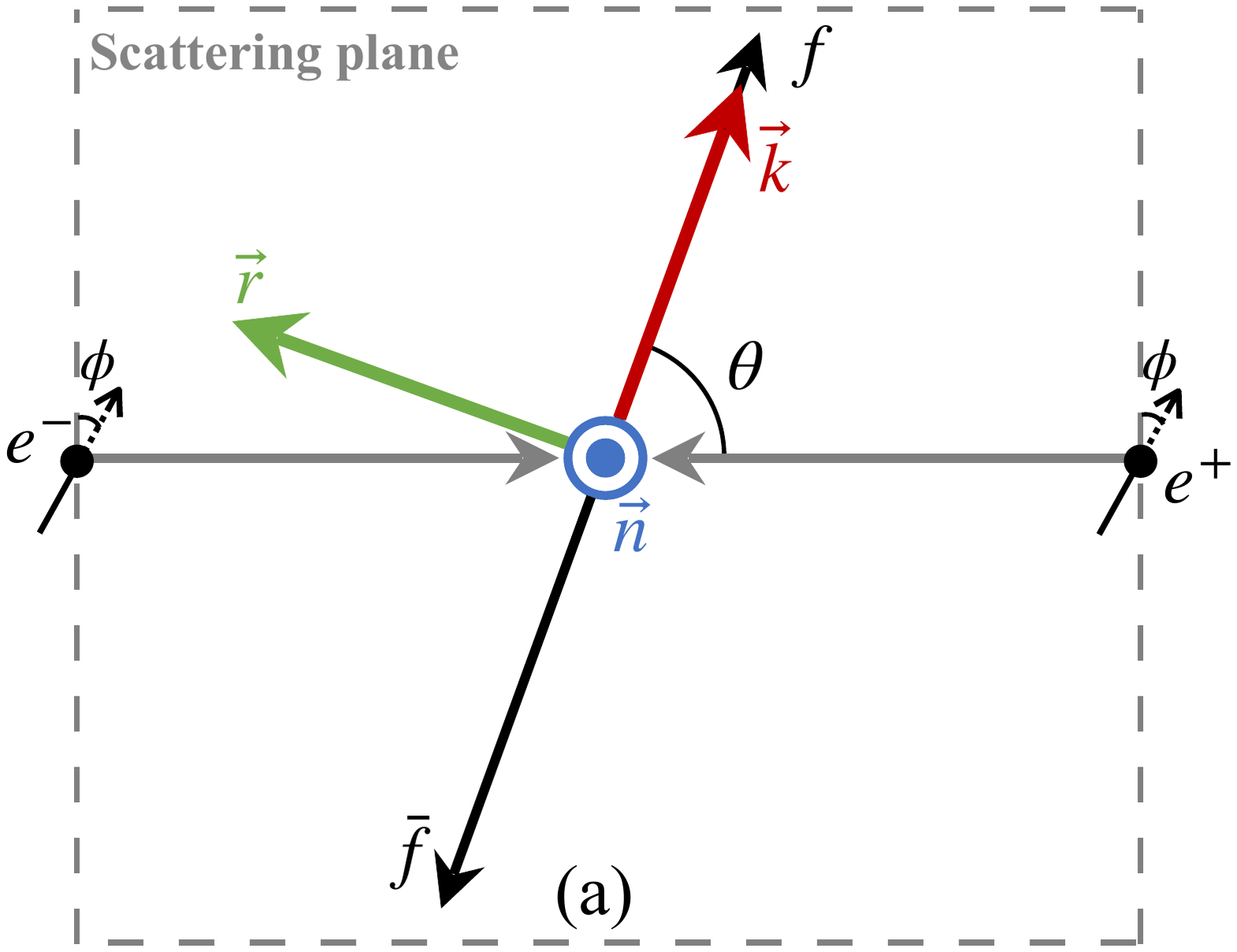}
    \includegraphics[width=0.4\linewidth]{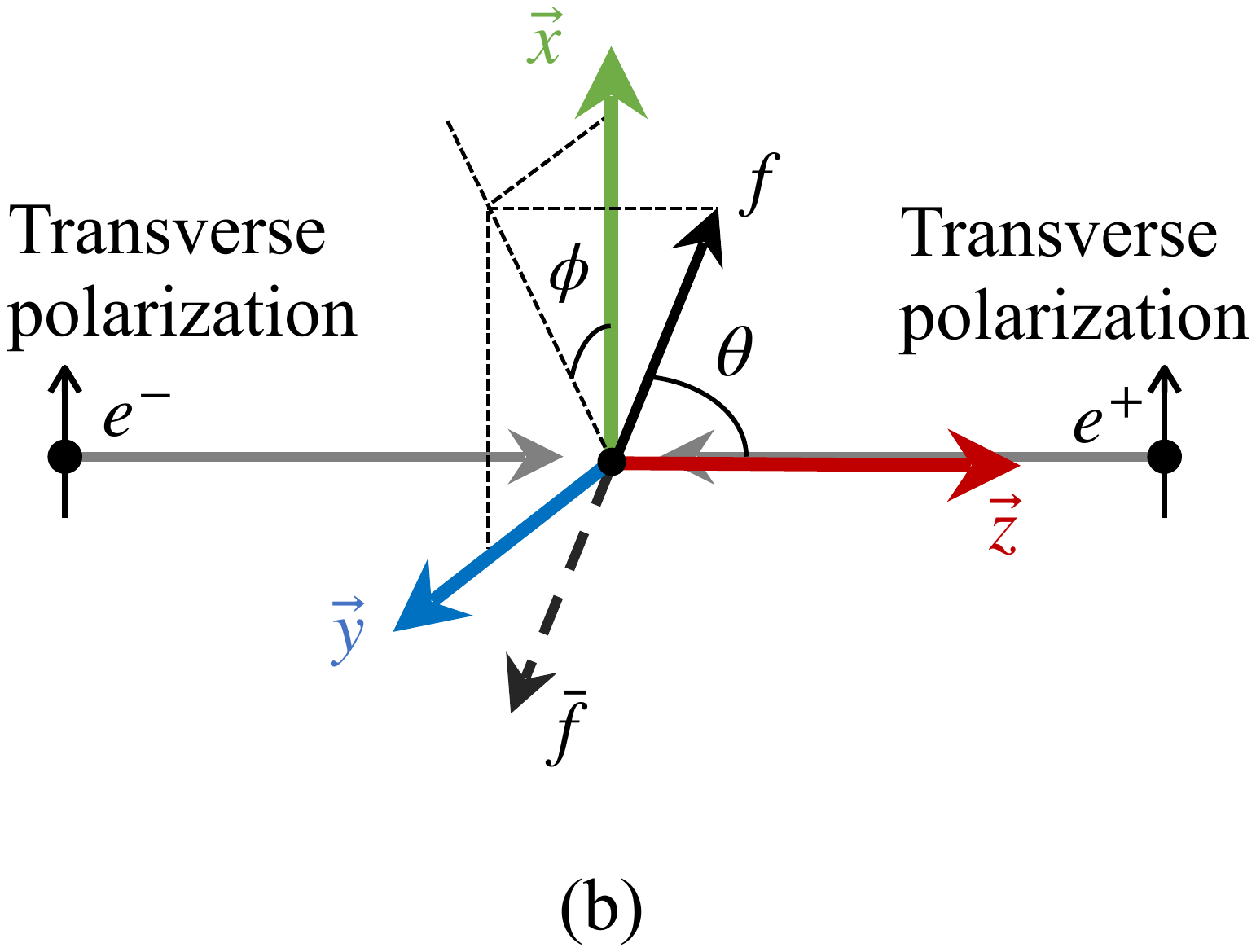}

	\caption{Different orthonormal basis in the c.m.~frame: (a) Helicity basis ($\hat{r},\hat{n},\hat{k}$) that varies with scattering angle, (b) Fixed beam basis ($\hat{x},\hat{y},\hat{z}$) that is the same for all scattering events. }
	\label{fig:basis}
\end{figure}

The helicity basis is defined with respect to the scattering plane, which usually leads to a simpler analytical expression for the amplitudes. In the rest of this work, we perform most of our calculations in the helicity basis, and $\ket{\uparrow}$ {($\ket{\downarrow}$)} with the omitted subscript denotes the $\ket{\uparrow_k}$  {($\ket{\downarrow_k}$)} in the helicity basis.
However, when the initial states are transversely polarized, the polarization vectors introduce an azimuthal reference direction, and therefore the scattering kinematics are no longer invariant under azimuthal rotation. In this case, the non-trivial $\phi$ dependence of the amplitudes must be taken into account.  It may also be useful to choose a fixed basis independent of the scattering kinematics, such as the fixed basis $\{\hat{x},\hat{y},\hat{z}\}$ shown in Fig.~\ref{fig:basis}(b), where $\hat{z}$ points along the initial $e^-$ beam, and $\hat{x}$ is the transverse polarization direction of the initial beams. 
In addition, different basis choices result in different angular averaged states, which leads to the concept of ``fictitious states''. We will comment on the implications in Section \ref{section:summary}.

\section{Pure QED Processes in Polarized \texorpdfstring{$e^-e^+$}{e- e+} Collisions}
\label{section:toymodel}

At electron-positron colliders, fermion pairs (except $e^-e^+$ and $\nu_e\bar\nu_e$) are predominantly produced from $s$-channel processes at the leading order.
To investigate how the entanglement of the final state behaves with transversely polarized beams, it is illustrative to start with purely QED interactions with only vector couplings of the photon mediator, which is the case for light fermion pair production below the $Z$-boson threshold.

\subsection{Entanglement with unpolarized and transversely polarized beams}
We first perform a general calculation of the final state density matrix using Eq.~\eqref{eq:mixed_state}.
In the massless limit, only $e^-e^+$ with opposite helicity can annihilate into a (virtual) vector boson, \textit{i.e.}, there are only $e_L^-e_R^+\to f_{\alpha}\bar{f}_{\bar\alpha}$ and  $e_R^-e_L^+\to f_{\alpha}\bar{f}_{\bar\alpha}$ processes.  
Their amplitudes, $\mathcal{M}^{LR}_{\alpha\bar\alpha}$ and $\mathcal{M}^{RL}_{\alpha\bar\alpha}$, are given by
\begin{equation}
    \mathcal{M}^{LR/RL}_{\alpha\bar{\alpha}}=\frac{e^2 Q_e Q_f}{s}J^{L/R}_{{\mathrm{in}},\mu}~(J^{\mu}_{{\mathrm{out}},V})_{\alpha\bar{\alpha}},
    \label{eq:amplitude_QED}
\end{equation}
where
\begin{align} \label{eq:JLJR}
J^{L}_{{\mathrm{in}},\mu}=\sqrt{s}\left(0,1,-i,0\right),\quad J^{R}_{{\mathrm{in}},\mu}=\sqrt{s}\left(0,-1,-i,0\right),
\end{align}
are the currents of different initial state configurations, $(J^{\nu}_{{\mathrm{out}},V})_{\alpha\bar{\alpha}}=\bar{u}_\alpha\gamma^\nu v_\alpha$ is the final state vector current, $Q_e=-1$ is the electron charge, and $Q_f$ is the charge of the fermion $f$. In the helicity basis, the scattering amplitudes are calculated as
\begin{subequations}\label{eq:amplitude_QED_explicit}
\begin{align}
    &\mathcal{M}^{LR}_{\alpha\bar\alpha}=e^2 Q_eQ_f \exp(-i \phi)
\begin{pmatrix}
  -(1-\cos\theta) \\
  -  \sin\theta /\gamma \\   
  - \sin\theta /\gamma \\
  - (1+\cos\theta ) \\
\end{pmatrix},\\
&\mathcal{M}^{RL}_{\alpha\bar\alpha}=e^2Q_eQ_f \exp(i \phi ) 
\begin{pmatrix}
 -  (1+\cos\theta )\\
 \sin\theta /\gamma \\
\sin\theta /\gamma \\
-(1-\cos\theta) \\
\end{pmatrix},
\end{align}
\end{subequations}
where $\beta\equiv \sqrt{1-4m_f^2/s}$ is the speed of $f$ in the c.m.~scattering frame, $\gamma=(1-\beta^2)^{-1/2}=\sqrt{s}/(2m_f)$ is the boost factor of $f$, and the spin index $\alpha\bar\alpha$ of $f\bar f$ is ordered as $\{\uparrow\uparrow,~\uparrow\downarrow, ~\downarrow\uparrow, ~\downarrow\downarrow\}$.

For unpolarized initial beams, the initial state density matrix of $e^-e^+$ is the identity matrix, and the explicit form of the final state density matrix is given in Eq.~(\ref{eq:rho_un}). Then, the spin correlation matrix from unpolarized scattering, ordered in the helicity basis $\{\hat r,\hat n,\hat k\}$,  is \cite{Cheng:2024btk}
\begin{equation}
C^{(\rm un)}=\left(
\begin{array}{ccc}
 \tfrac{2\sin^2\theta}{2-\beta ^2 \sin^2\theta}-\tfrac{\beta ^2 \sin^2\theta}{2-\beta ^2 \sin^2\theta} & 0 & -\tfrac{2\sin\theta \cos\theta \sqrt{1-\beta ^2}}{2-\beta ^2 \sin^2\theta} \\
 0 & -\tfrac{\beta ^2 \sin^2\theta}{2-\beta ^2 \sin^2\theta} & 0 \\
 -\tfrac{2\sin\theta \cos\theta \sqrt{1-\beta ^2}}{2-\beta ^2 \sin^2\theta} & 0 & \tfrac{2\cos^2\theta}{2-\beta ^2 \sin^2\theta}+\tfrac{\beta^2\sin^2\theta}{2-\beta ^2 \sin^2\theta} \\
\end{array}
\right).
\label{eq:qed_un_C}
\end{equation}
The entanglement of $f\bar f$ produced from similar unpolarized scattering processes is well studied ~\cite{Afik:2022kwm,Cheng:2024btk,Cheng:2025cuv}. It is commonly known that maximal entanglement only occurs in the high $p_T$ limit ($\beta\to 1$ and $\theta\to\pi/2$), where the spin correlation matrix $C^{({\mathrm{un}})}\to{\rm diag}(1,-1,1)$, and the state becomes close to the Bell state $\ket{\Psi}_{\hat{n}}$.

When both the $e^+$ and $e^-$ beams are transversely polarized along the $x$ direction, the density operator of the $e^\pm$ beam can be written as
\begin{subequations}\label{eq:rhoxab}
\begin{align} \label{eq:rhoxa}
\hat\rho^{e^\pm}&=B_x^{e^\pm}\left(\frac{\ket{R}+\ket{L}}{\sqrt2}\right)\left(\frac{\bra{R}+\bra{L}}{\sqrt2}\right)+(1-B_x^{e^\pm}) \frac{\hat{\mathbb{I}}}{2}\\
\label{eq:rhoxb}
&=\frac{\hat{\mathbb{I}}+B^{e^\pm}_x\hat{\sigma}_x}{2},
\end{align}
\end{subequations}
where the first term in the first line is the pure eigenstate of the $x$-component of the spin operator, and the identity operator $\hat{\mathbb{I}}/2$ in the first line is a completely unpolarized ``white noise''.
The mixture probability of the pure eigenstate of $\sigma_x$ in Eq.~\eqref{eq:rhoxa}, namely $B^{e^\pm}_x$, is the degree of polarization.

Starting with $B^{e^\pm}_x=100\%$ and plugging $\rho_{e^\pm}$ into Eq.~(\ref{eq:mixed_state}), we can obtain the explicit form of the final state density matrix, and the spin correlation matrix on the helicity basis is 
\begin{equation}
\label{eq:Cijxx}
C^{(xx)}=\frac{1}{M}\left(
\begin{array}{ccc}
M-2\sin^2\phi \cos^2\theta & -\cos\theta  \sin2\phi & -\frac{1}{\gamma } \sin2\theta \sin^2\phi \\
-\cos\theta \sin2\phi &M-2\cos^2\phi & -\frac{1}{\gamma }  \sin\theta  \sin2\phi  \\
-\frac{1}{\gamma } \sin2\theta \sin^2\phi & -\frac{1}{\gamma }\sin\theta \sin 2\phi &~~ M-\frac{2}{\gamma^2}\sin^2\phi\sin^2\theta  \\
\end{array}
\right)
\end{equation}
where the denominator $M$ is 
\begin{equation}
    M=1-\beta^2\sin^2\theta\sin^2\phi
\end{equation}
and the net polarization $B_i^{a/b}$ of the final state $f/\bar{f}$ is zero. 

The eigenvectors associated with the spin correlation matrix $C^{(xx)}$ are 
\begin{subequations}\label{eq:QEDdiagonalbasis}
\begin{align}
    \hat{e}_{1}&\propto \left(\cos\theta \cos\phi,~-(1-\beta^2\sin^2\theta)\sin\phi,~\frac{\sin\theta \cos\phi}{\gamma}\right),\\
    \hat{e}_{2}&\propto \left(\cos\theta \sin\phi, ~\cos\phi,~\frac{\sin\theta \sin\phi}{\gamma}\right), \label{eq:e2}\\
    \hat{e}_{3}&\propto \left(-\frac{\sin\theta}{\gamma},~0,~\cos\theta\right),
\end{align}
\end{subequations}
and the corresponding eigenvalues of $\{\hat e_1,\hat e_2, \hat e_3\}$ are $\{1, -1, 1\},$ respectively.
Therefore, the spin correlation matrix of $f\bar f$ can be decomposed as $C^{xx}=\mathbb{I}_3- 2\hat{e}_2\otimes \hat{e}_2$.
 Compared with the Bell states shown in Eq.~\eqref{eq:Bell_state}, we find that the state is a spin triplet $\ket{J=\mathbf{1},(\vec{J}\cdot\vec e_2)=0}$ state along $\hat e_2$, 
\begin{equation}\label{eq:QED_BellStateVector}
    \ket{\Psi}_{\hat e_2} = \frac{\ket{\uparrow_{\hat{e}_2}\downarrow_{\hat{e}_2}}+\ket{\downarrow_{\hat{e}_2}\uparrow_{\hat{e}_2}}}{\sqrt{2}},
\end{equation}
where the axis $\hat e_2$ depends on the momentum direction of the final-state fermion as shown in Eq.~\eqref{eq:e2}. 
The eigenvectors $\{\hat e_1,\hat e_2, \hat e_3\}$ also form a set of orthogonal basis, known as the diagonal basis~\cite{Cheng:2024btk}, which is the optimal basis choice when averaging the spin state over the phase space.

There are two major differences between the spin state produced by the scattering of unpolarized beams and the transversely polarized beams. Firstly and most remarkably, transverse polarization significantly increases the entanglement, and 100\% transversely polarized beams would result in maximally entangled final states regardless of scattering energy and angle. Secondly, in the presence of transverse polarization, the final spin state has a physical azimuthal angle dependence as shown in Eq.~\eqref{eq:Cijxx}.  The transverse polarization introduces a reference direction regardless of scattering kinematics as shown in Fig.~\ref{fig:basis}(b), so that the scattering process does not exhibit rotational symmetry in the azimuthal angle. This also usually makes the helicity basis less convenient for describing the spin state.
For example, in the threshold limit ($\beta\to0,\ \gamma\to1$), the spin state has an azimuthal angle-independent form in the fixed basis instead of the helicity basis. We have
\begin{equation}
    \hat{e}_2\propto \hat{y}+\hat k\left(\frac{1}{\gamma}-1\right)\sin\theta\sin\phi\to\hat y,
\end{equation}
where the unit directions $\hat{r},\hat{n},\hat{k}$ and $\hat{y}$ are shown in Fig.~\ref{fig:basis}. The final state is $\ket{\Psi}_{\hat{y}}$, with a negative spin correlation along the $\hat{y}$ direction and a positive spin correlation along the $\hat{x}$ and $\hat{z}$ directions.

The concurrence of $f\bar f$ produced from QED processes with unpolarized and 100\% transversely polarized beams is shown,  respectively, in Fig.~\ref{fig:QEDConcurrence}(a) and Fig.~\ref{fig:QEDConcurrence}(d) in the plane of the scattering angle $\cos\theta$ and the speed $\beta$ in the $f\bar f$ c.m.~frame. The color code and dashed contour lines indicate the values of the concurrence and thus the level of entanglement. 
As discussed above, with unpolarized beams, the entanglement of $f\bar f$ is usually weak unless the scattering process occurs ultra-relativistically in the central region as the $f\bar f$ state approaches an entangled triplet state. However, with 100\% transversely polarized beams, the final states are always maximally entangled as a spin triplet at any phase space point, while the triplet configuration can be different event-by-event as seen in Eq.~(\ref{eq:QED_BellStateVector}).

\begin{figure}[tb]
	\centering
    \includegraphics[width=\linewidth]{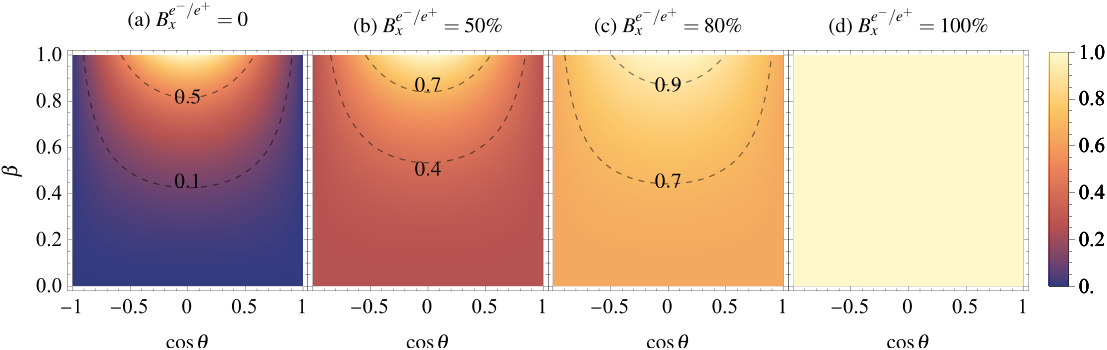}

	\caption{Concurrence of the spin density matrix for the QED process $e^-e^+ \to \gamma^*\to  f \bar f$ in the $\cos\theta$-$\beta$ plane in the c.m.~frame with $\phi=0$: (a) unpolarized, (b) 50\% transversely polarized, (c) 80\% transversely polarized, and (d) 100\% transversely polarized beams.}. 
	\label{fig:QEDConcurrence}
\end{figure}

More generally, for an arbitrary beam transverse polarization $|B_x^{e^\pm}|<1$, the final state $\rho^{f\bar f}$ can be written as a mixture
\begin{equation}\label{eq:generalPolarization}
    \rho^{f\bar f}_{\alpha\bar\alpha,\alpha'\bar\alpha'}=\omega^{\rm (un)}\rho_{\alpha\bar\alpha,\alpha'\bar\alpha'}^{\rm (un)}+\omega^{(xx)}\rho_{\alpha\bar\alpha,\alpha'\bar\alpha'}^{(xx)},
\end{equation}
where $\rho^{\rm (un)}$ and $\rho^{(xx)}$ are the normalized density matrices produced from the scattering of unpolarized and 100\%  transversely polarized beams, respectively. The weights of the mixture are
\begin{equation}
    \omega^{\rm (un)}=\frac{d\sigma^{\rm (un)}}{d\sigma^{f\bar f}}(1-B_x^{e^-}B_x^{e^+}),\quad \omega^{(xx)}=\frac{d\sigma^{(xx)}}{d\sigma^{f\bar f}}B_x^{e^-}B_x^{e^+}, \quad \omega^{\rm (un)}+\omega^{(xx)}=1,
\end{equation}
where $d\sigma^{\rm (un)}\ (d\sigma^{(xx)})$ is the differential cross section for the unpolarized case (100\% transversely polarized case), and 
$d\sigma^{f\bar f}=(1-B_x^{e^-}B_x^{e^+}) d\sigma^{\rm (un)} +B_x^{e^-}B_x^{e^+} d\sigma^{(xx)}$ is the differential cross section with partially transversely polarized beams.  The correlation matrix is of the same mixture of Eq.~\eqref{eq:qed_un_C} and Eq.~\eqref{eq:Cijxx}, and the concurrence can be calculated analytically as
\begin{equation}
    \mathscr{C}[\hat\rho]=\sqrt{1-\frac{4[1-(B_x^{e^-}B_x^{e^+})^2](1-\beta^2\sin^2\theta)}{(2-\beta^2\sin^2\theta+B_x^{e^-}B_x^{e^+}\beta^2\sin^2\theta\cos2\phi)^2}}.
\end{equation}
The explicit expressions for the density matrix and the spin correlation matrix as functions of $B_x^{e^\pm}$ are shown in Appendix~\ref{appendix:genneral_Cij}.

Consequently, the entanglement of the final state changes continuously as we increase the transverse polarization of the initial beam, as shown in Figs.~\ref{fig:QEDConcurrence}(b) and (c).
Additionally, to observe the unique feature induced by the transverse polarization, both initial beams have to be polarized.
As long as one beam is unpolarized, the transverse polarization of the other beam does not make any difference, and the final spin state results in the unpolarized scattering case.

\subsection{Maximal entanglement from the diagonal amplitudes}
\label{sub:diagonalBasis}
In addition to the general calculation in the previous subsection, for an intuitive understanding of how transverse polarization increases the entanglement between final states, it is also useful to study the special case of purely transversely polarized beams at the amplitude level.

With 100\% transversely polarized beams, both the initial and the final states are pure states.
The spin state $\ket{e^-e^+}$ of the initial beams is given in Eq.~\eqref{eq:statevector_intro}, where the state vector is $\psi^{ee}_{\lambda\bar\lambda}\propto (1,1,1,1)^{\rm T}$, and the spin state vector of the $f\bar f$ final state is obtained by multiplying the transition matrix with the initial state spin state vector,
\begin{equation}
    \ket{f\bar f} = \hat{\mathcal{T}}\ket{e^-e^+}.
\end{equation}
The transition matrix in the spin space of the scattering process is
\begin{equation}
\renewcommand{\arraystretch}{1.5}
    \mathcal{T}_{\alpha\bar\alpha}^{\lambda\bar\lambda}=\begin{pmatrix}
        \mathcal{M}^{RL}_{\uparrow\uparrow} &0 &0 & \mathcal{M}^{LR}_{\uparrow\uparrow} \\
        \mathcal{M}^{RL}_{\uparrow\downarrow} &0 &0 & \mathcal{M}^{LR}_{\uparrow\downarrow} \\
        \mathcal{M}^{RL}_{\downarrow\uparrow} &0 &0 & \mathcal{M}^{LR}_{\downarrow\uparrow} \\
        \mathcal{M}^{RL}_{\downarrow\downarrow} &0 &0 & \mathcal{M}^{LR}_{\downarrow\downarrow} \\
    \end{pmatrix},
\end{equation}
where the explicit forms of the amplitudes are given in Eq.~\eqref{eq:amplitude_QED_explicit}.

\begin{figure}[tb]
    \centering
    \includegraphics[width=0.4\linewidth]{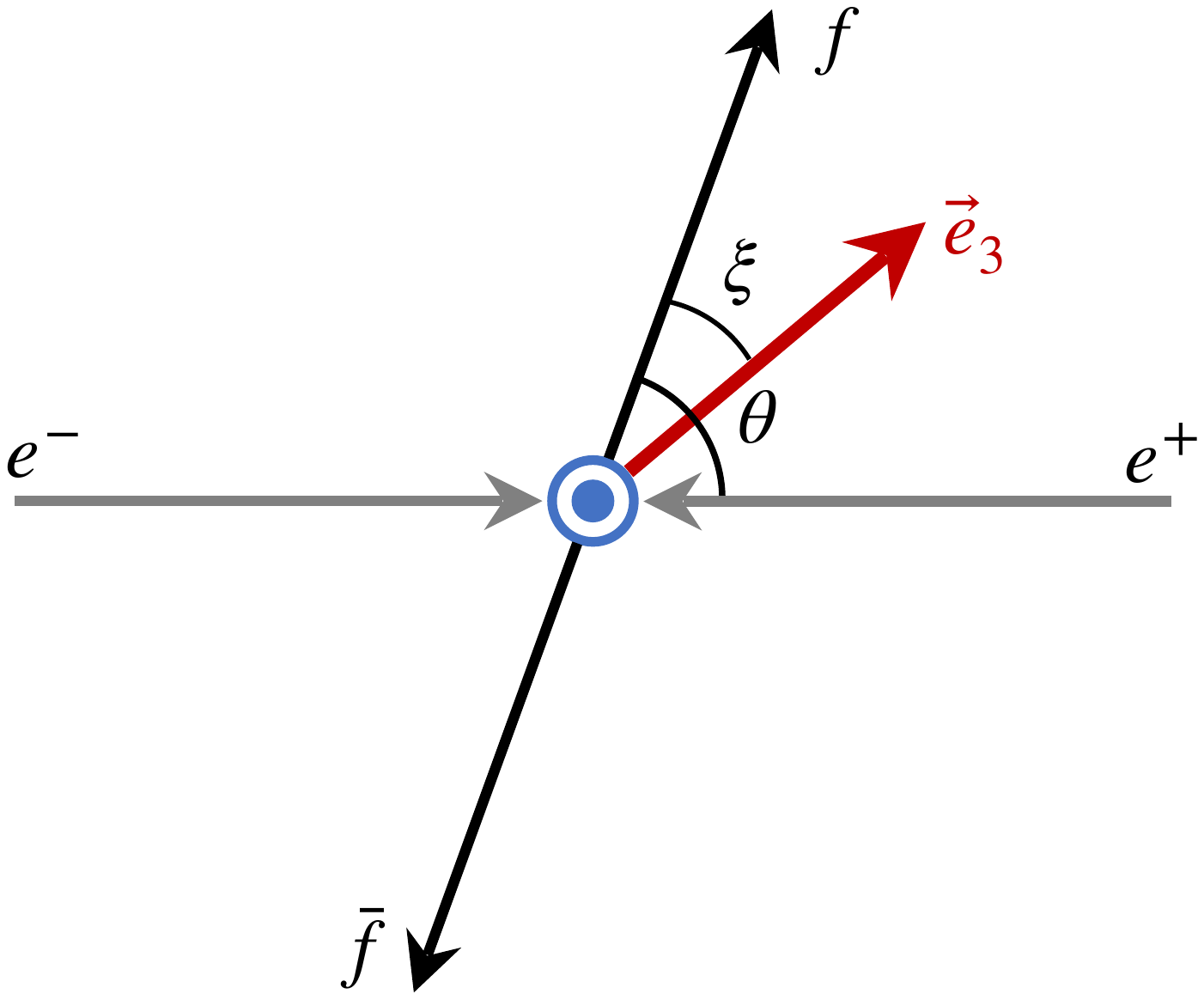}
    \caption{The diagonal basis: the spin quantization axis $\hat{e}_3$ under which the $f$ and $\bar f$ have the same spin projection.}
    \label{fig:diagonalBasis}
\end{figure}

One can prove that the final state vector $\ket{f\bar f}$ is maximally entangled in the ``diagonal basis"~\cite{Parke:1996pr,Mahlon:1997uc}.
For an $s$-channel process with only vector current interactions, it is known that there exists a basis choice under which only the amplitude for like-spin final states is non-zero~\cite{Parke:1996pr,Mahlon:1997uc}.\footnote{ In Refs.~\cite{Parke:1996pr,Mahlon:1997uc} the spin quantization axis for $f$ and $\bar f$ are chosen to be opposite, therefore the basis is referred to as ``off-diagonal basis". Same as Ref.~\cite{Cheng:2023qmz,Cheng:2024btk}, we choose the same spin quantization axis for $f$ and $\bar f$, therefore the helicity amplitude $\mathcal{M}^{\uparrow\uparrow/\downarrow\downarrow}_{\alpha\bar\alpha}$ is diagonal in this basis, and we refer to it as ``diagonal basis".}
Namely, for both left- and right-handed initial states, there are only two non-zero amplitudes $\mathcal{M}_{\uparrow\uparrow}$ and $\mathcal{M}_{\downarrow\downarrow}$.
The spin quantization axis $\hat{e}_3$ is shown in Fig.~\ref{fig:diagonalBasis} where $\tan\xi=\tan\theta/\gamma$~\cite{Mahlon:1997uc}.
In this basis, the transition matrix is rewritten as
\begin{equation}\label{eq:transition_diagonalbasis}
    \mathcal{T}_{\alpha\bar\alpha}^{\lambda\bar\lambda}\propto \begin{pmatrix}
        e^{i\phi} A~ &0 &0 & e^{-i\phi} B \\
        0&0&0&0\\
        0&0&0&0\\
        e^{i\phi} B~ &0 &0 & e^{-i\phi} A \\
    \end{pmatrix},
\end{equation}
where
\begin{align}
  A= -\frac{\sin\theta\sin\xi}{\gamma}-\cos\theta\cos\xi-1,\qquad 
  B=\frac{\sin\theta\sin\xi}{\gamma}+\cos\theta\cos\xi-1.
\end{align}
Multiplying the transition matrix $\mathcal{T}_{\alpha\bar\alpha}^{\lambda\bar\lambda}$ from Eq.~\eqref{eq:transition_diagonalbasis} with the state vector $\psi^{ee}_{\lambda\bar\lambda}\propto (1,1,1,1)^{\rm T}$, we obtain the final state vector $\psi^{f\bar f}_{\alpha\bar\alpha}\propto (x,0,0,x^*)^{\rm T}$ with $x\equiv A e^{i\phi}+B e^{-i\phi}$, and the spin state of $f\bar f$ in the diagonal basis is
\begin{equation}\label{eq:QEDBellState_Proof}
  \ket{f\bar f} = \frac{1}{\sqrt{2}} (e^{i \arg(x)}\ket{\uparrow_{\hat{e}_{3}}\uparrow_{\hat{e}_{3}}} +  e^{-i \arg(x)}\ket{\downarrow_{\hat{e}_3}\downarrow_{\hat{e}_3}}),
\end{equation}
which is a maximally entangled spin triplet.  For example, in the massless limit with $\beta\to 1$, $\hat{e}_{3}=\hat{k}$ and the triplet state configuration is $\ket{\Psi}_{\hat{n}}$ at $\phi=0$.

In fact, as long as the $e^-$ and $e^+$ beams are 100\% transversely polarized, the final state $f\bar f$ produced from the QED process is always maximally entangled regardless of the directions of the polarization vector of the $e^\pm$ beams. Similarly to Eq.~\eqref{eq:statevector_intro}, if the $e^-$ beam is transversely polarized along the $\phi_1$ direction and the $e^+$ beam is 100\% transversely polarized along the $\phi_2$ direction, then their spin state is
\begin{align}
    &\frac{1}{\sqrt{2}}(\ket{R}+e^{i\phi_1}\ket{L})\otimes
    \frac{1}{\sqrt{2}}(\ket{L}+e^{i\phi_2}\ket{R}) \nonumber\\
    =& \frac{1}{2}(\ket{RL}+e^{i\phi_2}\ket{RR}+e^{i\phi_1}\ket{LL}+e^{i(\phi_1+\phi_2)}\ket{LR}).
\end{align}
Then the spin state vector of $f\bar f$ in the diagonal basis is $\psi^{f\bar f}_{\alpha\bar\alpha}\propto (x',0,0,-x'^*)$ where $x'\equiv A e^{-i\phi'}-B e^{i\phi'}$ with $\phi'=\phi+\frac{\phi_1+\phi_2}{2}$.  Therefore, the final state is still a maximally entangled spin triplet
\begin{equation}
  \ket{f\bar f} = \frac{1}{\sqrt{2}} (e^{i \arg(x')}\ket{\uparrow_{\hat{e}_3}\uparrow_{\hat{e}_3}} +  e^{-i \arg(x')}\ket{\downarrow_{\hat{e}_3}\downarrow_{\hat{e}_3}}).
\end{equation}
For simplicity and without loss of generality, in this work we only consider the case where the $e^+$ and $e^-$ beams are polarized along the same transverse direction.

\subsection{Quantum magic}

We also consider quantum magic for a comparative description of the quantum state produced.
In the language of quantum simulation, a stabilizer state is a quantum state that does not possess an advantage in a quantum simulation.
A convenient way to quantify whether a general state $\rho$ is stabilizer or not is through its \emph{Pauli spectrum}, the collection of expectation values $\{\mathrm{Tr}[P\rho], \, P\in{\cal P}_n\}$, where the Pauli string is defined as $\big\{ {\cal P}_n = P_1 \otimes P_2 \otimes \ldots \otimes P_n \big| 
   P_i \in \{\mathbbm{1},\sigma_1,\sigma_2,\sigma_3\}\,\, \big\}$. The subgroup of unitary operators that map Pauli strings to each other in conjugation forms the \emph{Clifford group}. Applying the quantum gates in the Clifford group on a reference state produces the set of \emph{stabilizer states} that admit efficient classical simulation.   

To quantify deviations from the stabilizer set, one may introduce the stabilizer Rényi entropies~\cite{Leone:2021rzd}   
\begin{equation}
  \mathcal{M}_k(\rho)=\frac{1}{1-k}\log_2\!\left(\frac{\sum_{P\in{\cal P}_n} (\mathrm{Tr}[P\rho])^{2k}}
  {\sum_{P\in{\cal P}_n} (\mathrm{Tr}[P\rho])^2}\right),
  \label{Mqdef}
\end{equation}
defined for integer $k\geq 2$. 
In practice, the $k=2$ case, known as the \emph{second stabilizer Rényi entropy} (SSRE), is widely used as a robust probe of magic.
For a qubit pair parametrized as Eq.~\eqref{eq:rhoCij}, the second stabilizer Rényi entropy ($\mathcal{M}_2$) is given by~\cite{White:2024nuc}
\begin{equation}\label{eq:magic}
    \mathcal{M}_2 = - \log_2 \left( \frac{1+\sum_{i}(B_{i}^{a})^4 +\sum_{j}(B_{j}^{b})^4 +\sum_{i,j}C_{ij}^4}{1+\sum_{i}(B_{i}^{a})^2 +\sum_{j}(B_{j}^{b})^2 +\sum_{i,j}C_{ij}^2} \right).
\end{equation}
The measure $\mathcal{M}_2$ vanishes for stabilizer states and increases as the state departs from the stabilizer set.

\begin{figure}
    \centering
    \includegraphics[width=0.65\linewidth]{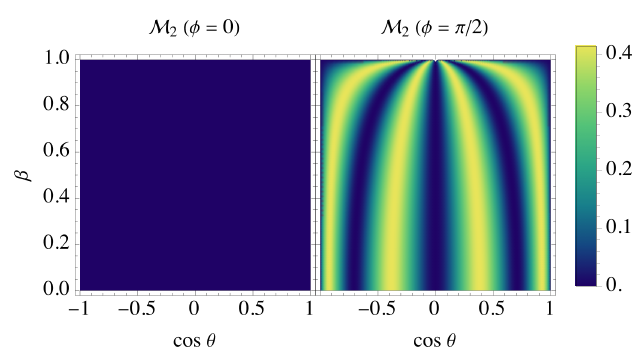} \caption{Quantum magic (SSRE) for the QED process $e^-e^+ \to \gamma^*\to  f \bar f$ in the $\cos\theta$-$\beta$ plane  with 100\% transversely polarized beams, for (a) $\phi=0$ and (b) $\phi=\pi/2$.}
    \label{fig:QEDmagic}
\end{figure}

Unlike the quantum entanglement that is basis independent, quantum magic is defined to be basis dependent as seen from Eq.~\eqref{eq:magic}.  For maximally entangled Bell states given in Eq.~\eqref{eq:Bell_state}, they are stabilizer states that can be produced directly by acting Clifford gates on a reference state $\ket{00}$. However, if the Pauli chain is not chosen to be the same basis as Eq.~\eqref{eq:Bell_state}, then the Bell state can have non-zero magic.
In Fig.~\ref{fig:QEDmagic}, we show the quantum magic of a fermion pair produced from the QED process, where we choose the helicity basis to define the Pauli chain.  Although the state is a maximally entangled Bell state at any scattering angle, its spin configuration varies with phase space.
In the helicity basis, when examining the events that are coplanar with the beam polarization direction ($\phi=0$), the state is a stabilizer with zero magic as seen in Fig.~\ref{fig:QEDmagic}(a) in the $\cos\theta$-$\beta$ plane.
When examining the events produced in other phase space with nonzero $\phi$, the spin state could be a non-stabilizer state in the helicity basis, as shown in  Fig.~\ref{fig:QEDmagic}(b) with $\phi=\pi/2$. These features can be seen explicitly from Eq.~(\ref{eq:Cijxx}).

\section{Electro-weak Processes in Polarized \texorpdfstring{$e^-e^+$}{e- e+} Collisions}
\label{section:Spin_correlation}
At a high-energy $e^- e^+$ collider, fermion pairs $f\bar f$ are produced through $s$-channel processes via both $\gamma$ and $Z$ exchanges. After including the $Z$ boson contribution, both vector and axial vector currents contribute to the scattering amplitude, Eq.~\eqref{eq:amplitude_QED} is now generally parametrized as
\begin{align}\label{eq:EWamplitudes}
    \mathcal{M}^{LR/RL}_{\alpha\bar{\alpha}}=\frac{e^2}{s}J^{L/R}_{\rm  in,\mu}\left(f_V^{L/R}(J^{\mu}_{\mathrm{out},V})_{\alpha\bar{\alpha}}+f_A^{L/R}(J^{\mu}_{\mathrm{out},A})_{\alpha\bar{\alpha}}\right).
\end{align}
Here, $J^{L/R}_{\rm in, \mu} =\frac{\sqrt{s}}{2} (0,\pm 1,-i,0)$ is the same initial state current as in Eq.~\eqref{eq:JLJR}, $(J^{\nu}_{\mathrm{out},V})_{\alpha\bar{\alpha}}=\bar{u}_{\alpha}\gamma^\nu v_{\bar\alpha}$ and   $(J^{\nu}_{\mathrm{out},A})_{\alpha\bar{\alpha}}=\bar{u}_{\alpha}\gamma^\nu\gamma^5 v_{\bar\alpha}$ are the vector and axial vector currents of the final states, with the interaction strengths $f_V$ and $f_A$, respectively. The interaction strengths are
\begin{subequations}
    \label{eq:coefficients}
\begin{align}
    &f_V^L=Q_eQ_f+\frac{g^e_L(g^f_R+g^f_L)}{2c_W^2s_W^2}\frac{s}{s-m_Z^2},\\
    &f_A^L=\frac{g^e_L(g^f_R-g^f_L)}{2c_W^2s_W^2}\frac{s}{s-m_Z^2},\\
    &f_V^R=Q_eQ_f+\frac{g_R^e(g_R^f+g_L^f)}{2c_W^2 s_W^2}\frac{s}{s-m_Z^2},\\
    &f_A^R=\frac{g^e_R(g^f_R-g^f_L)}{2c_W^2 s_W^2}\frac{s}{s-m_Z^2}.
\end{align}
\end{subequations}
The chiral couplings of fermions with the $Z$  boson are $g_L^f= I_f^3-Q_f s_W^2$ and $g_R^f=-Q_f s_W^2$.
The convention is $Q_e=-1$ and $I_e^3=-1/2$. 

The quantum behavior of the final state $f\bar f$ can be drastically different depending on the chiral interactions and kinematics. This is governed by the relative contributions of the vector and axial vector currents $f^{L/R}_A/f^{L/R}_V$. To show those features, we choose to present the representative processes  
\begin{equation}
e^-e^+  \to t\bar t,\ \  \tau^-\tau^+ \ \ {\rm and}\ \ b \bar b.  
\end{equation}

\subsection{\texorpdfstring{$e^-e^+ \to t \bar t$}{e- e+  ->  t tbar}}

We begin with the $e^-e^+\to t\bar t$ process, which is well above the $Z$ boson threshold.
This process is dominated by the vector interactions since the vector coupling constant is larger than the axial vector coupling constant. As such, the ratio between the vector and the axial vector coupling in the $e^-e^+\to t\bar t$ process has little dependence on energy, as shown in Fig.~\ref{fig:fVfA}, and the energy dependence of the spin state $t\bar t$ is simply the result of a different speed $\beta$.

\begin{figure}[tb]
	\centering
    \includegraphics[width=0.45\linewidth]{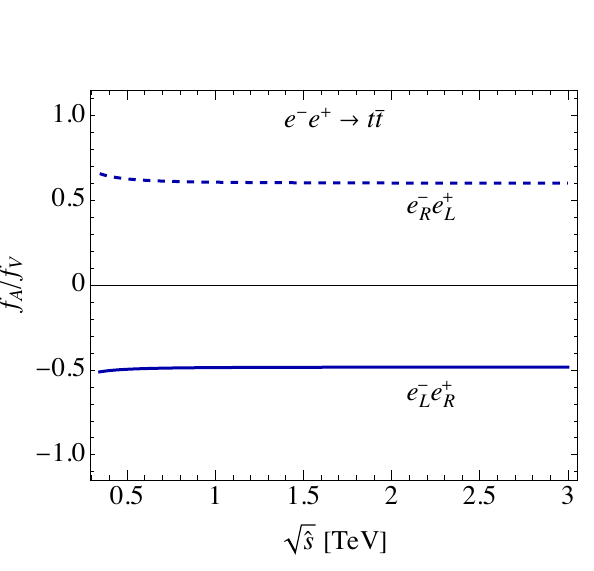}

	\caption{Ratio between the axial vector and vector interaction strengths $f^L_A/f^L_V$ and $f^R_A/f^R_V$ for the EW process $e^-e^+\to t\bar{t}$.
    }
	\label{fig:fVfA}
\end{figure}

\begin{figure}[tb]
	\centering
    \includegraphics[width=\linewidth]{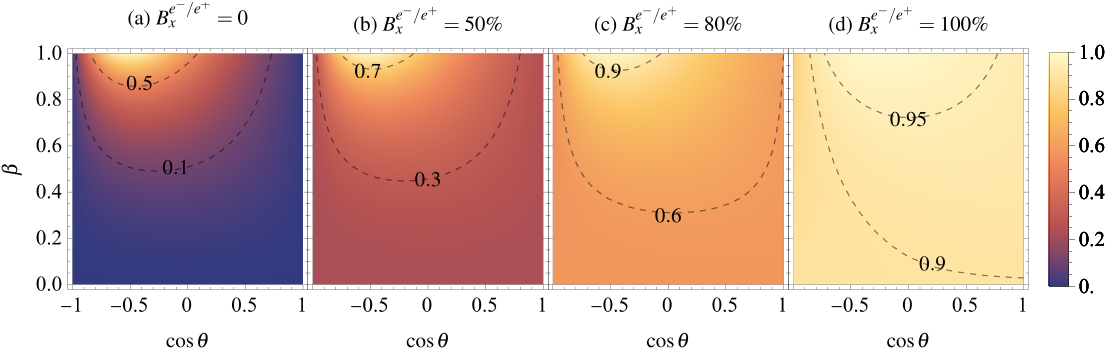}

	\caption{Concurrence of the spin density matrix for the EW process $e^-e^+\to t\bar{t}$ in the $\cos\theta$-$\beta$ plane in the $t\bar{t}$ c.m.~frame with $\phi=0$: (a) unpolarized, (b) 50\% transversely polarized, (c) 80\% transversely polarized, and (d) 100\% transversely polarized beams.}. 
	\label{fig:xxConcurrence}
\end{figure}

Since the vector current interaction always dominates, the quantum properties of the $e^-e^+\to t\bar t$ process are similar to the QED process.
Figure~\ref{fig:xxConcurrence} illustrates the entanglement of $t\bar t$ with different collision energies, scattering angles, and transverse polarizations of beams.
As expected, the overall behavior is similar to the QED process in Fig.~\ref{fig:QEDConcurrence}.  
The slight difference from the QED process arises from the contributions of the axial current.  The axial current is small and does not change the result qualitatively. It leads to chiral couplings and a preference for left- and right-handed final states, which usually reduces the entanglement between final state particles. 
In addition, the contribution of the axial vector current also violates parity and introduces forward-backward asymmetry in the final state.

In the unpolarized case, the concurrence in most phase space is small unless in the ultra-relativistic region, as in the QED process. The difference between the QED case is that the central scattering region ($\cos\theta=0$) does not yield maximal entanglement.
Given a scattering energy, the concurrence maximizes at $\cos\theta \approx-\beta|f_A/f_V|$ due to the contribution of axial currents~\cite{Cheng:2024btk}. 

When both initial beams are 100\% transversely polarized along the direction $\hat x$, the final states are entangled in the entire phase space as shown in Fig.~\ref{fig:xxConcurrence}(d), which is also similar to the QED process.  However, due to the involvement of the axial vector current, the helicity amplitudes of $e_L^-e_R^+\to t\bar t$ and $e_R^-e_L^+\to t\bar t$ cannot be diagonalized simultaneously as the QED case~\cite{Parke:1996pr}, and the transition matrix cannot be written in the form of Eq.~\eqref{eq:transition_diagonalbasis}. Then the $t\bar t$ state is not exactly a Bell state as in Eq.~\eqref{eq:QEDBellState_Proof}, and the values of concurrence are smaller than 1.

Figures~\ref{fig:xxConcurrence}(b) and (c) also show the entanglement of $t\bar t$ when the initial beams are partially transversely polarized.
As given in Eq.~\eqref{eq:generalPolarization}, the final state is a mixture of the unpolarized case and the 100\% polarized case, and the entanglement of the final state changes continuously with the transverse polarization of the initial beams.

\subsection{Light fermion pairs in \texorpdfstring{$e^-e^+$}{e- e+} collisions}
We next consider light fermion pairs such as $\tau^-\tau^+$ and $b\bar b$ that can be produced either above or below the $Z$ threshold. The helicity amplitudes have the same form as given in Eqs.~\eqref{eq:EWamplitudes} and~\eqref{eq:coefficients}, except that when approaching the $Z$-pole the Breit-Wigner propagator should take the form $(s-m_Z^2)^{-1} \to \left(s-m_Z^2+im_Z \Gamma_Z(s)\right)^{-1}\approx \left(s-m_Z^2+is\Gamma_Z/m_Z\right)^{-1}$~\cite{Bardin:1988xt,Borrelli:1989bd}.

In contrast to the $e^-e^+\to t\bar{t}$ process where the vector coupling is always dominant, the chiral structures of the $\tau$ lepton and $b$ quark have a significant energy dependence when crossing the $Z$ threshold.
In Fig~\ref{fig:fVfAtautau}, we show the ratio between the contributions from the vector and axial vector of the final state.
In particular, we see that the axial current coupling generally dominates around the $Z$ threshold. 
Therefore, unlike the QED and $e^-e^+\to t\bar t$ processes, the entanglement of $\tau^-\tau^+$ and $b\bar b$ at different energies is mostly determined by the axial/vector coupling strengths instead of the kinematic variable $\beta$.

\begin{figure}
	\centering
    \includegraphics[width=0.47\linewidth]{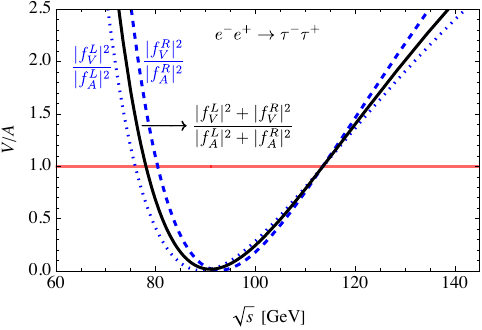}
    \includegraphics[width=0.48\linewidth]{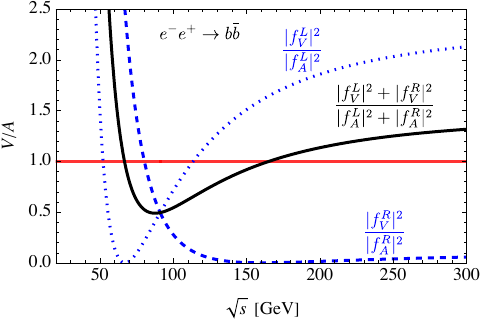}

	\caption{The energy dependence of the vector/axial current interaction ratio for  (a)  $e^-e^+\to \tau^-\tau^+$ and (b)  $e^-e^+\to b \bar b$.
    }
	\label{fig:fVfAtautau}
\end{figure}

To show how the energy dependence of $f_{V/A}$ affects the quantum state, it is convenient to consider the massless limit, where the amplitudes $\mathcal{M}_{\uparrow\downarrow/\downarrow\uparrow}$ vanish in the helicity basis.\footnote{The analysis in the massless limit captures the overall picture. Nevertheless, in our later numerical results of Figs.~\ref{fig:tauConcurrences} and~\ref{fig:bbConcurrences}, the $\tau$ lepton mass ($m_\tau=1.777$  GeV) and the bottom quark mass ($m_b = 4.198$ GeV) are taken into account.  }
The transition matrix in the massless limit is
\begin{equation}\label{eq:transitionmatrixMasslessTau}
  \mathcal{T}_{\alpha\bar\alpha,\lambda\bar\lambda}= \begin{pmatrix}
        \mathcal{M}^{RL}_{\uparrow\uparrow} &0 &0 & \mathcal{M}^{LR}_{\uparrow\uparrow} \\
        0 &0 &0 & 0 \\
        0 &0 &0 & 0 \\
        \mathcal{M}^{RL}_{\downarrow\downarrow} &0 &0 & \mathcal{M}^{LR}_{\downarrow\downarrow} \\
    \end{pmatrix}=
    \begin{pmatrix}
       (f^R_V+f^R_A) \frac{1+\cos\theta}{2}e^{i\phi} &0 &0 &  (f^L_V+f^L_A)\frac{1-\cos\theta}{2}e^{-i\phi} \\
        0&0&0&0\\
        0&0&0&0\\
        (f^R_V-f^R_A) \frac{1-\cos\theta}{2}e^{i\phi} &0 &0 &  (f^L_V-f^L_A) \frac{1+\cos\theta}{2}e^{-i\phi} \\
    \end{pmatrix}.
\end{equation}
The \textit{unnormalized} spin states of $\tau^-\tau^+$ pair produced from left- and right-handed initial states are, respectively,
\begin{subequations}\label{eq:statevectorLR}
\begin{align}
   \ket{f\bar f}_L & =(f^L_V+f^L_A) \frac{1-\cos\theta}{2}e^{-i\phi}  \ket{\uparrow\uparrow} + (f^L_V-f^L_A) \frac{1+\cos\theta}{2}e^{-i\phi} \ket{\downarrow\downarrow} \\
    \ket{f\bar f}_R  &=(f^R_V+f^R_A) \frac{1+\cos\theta}{2}e^{i\phi}  \ket{\uparrow\uparrow} + (f^R_V-f^R_A) \frac{1-\cos\theta}{2}e^{i\phi} \ket{\downarrow\downarrow} 
\end{align}
\end{subequations}
The superposition or mixing of the above two states sensitively depends on the values of $f_V$ and $f_A$.
For both left- and right-handed initial states, when the vector current dominates ($|f_V|>|f_A|$), the fermion pair $\ket{f\bar f}$ state is close to the spin triplet $\ket{\Psi}_{\hat{n}}=(\ket{\uparrow\uparrow}+\ket{\downarrow\downarrow})/\sqrt{2}$; when the axial current dominates ($|f_A|>|f_V|$), the $\ket{f\bar f}$ state is close to a different triplet configuration $\ket{\Psi}_{\hat{r}}=(\ket{\uparrow\uparrow}-\ket{\downarrow\downarrow})/\sqrt{2}$.

\vspace{10pt}
\noindent\textbf{Unpolarized scattering}

In the case of unpolarized scattering, the initial beams are incoherent mixtures of left- and right-handed states.  Therefore, the final state is also a mixed state of the two states in Eq.~\eqref{eq:statevectorLR}, given by
\begin{equation}\label{eq:LRmixing}
    \rho_{f\bar f}\propto \ket{f\bar f}_L \bra{f \bar f}_L + \ket{f\bar f}_R \bra{f\bar f}_R
\end{equation}
If the vector or axial vector coupling is dominant for both initial state polarizations, then the spin configurations of $\ket{f\bar f}_R$ and  $\ket{f\bar f}_L$ are similar and their mixing is also a similar entangled state.
However, if the contributions from vector and axial current interactions are comparable, the off-diagonal terms, i.e., the $\ket{\uparrow\uparrow}\bra{\downarrow\downarrow}+h.c.$ terms in Eq.~\eqref{eq:LRmixing} from $\ket{f\bar f}_L$ and $\ket{f\bar f}_R$, cancel each other when
\begin{equation}\label{eq:VeqA}
    (f^L_V)^2+ (f^R_V)^2=
    (f^L_A)^2+ (f^R_A)^2,
\end{equation}
resulting in a separable state.  This could happen for both $\tau^-\tau^+$ and $b\bar b$ as shown in Fig.~\ref{fig:fVfAtautau}.

\vspace{10pt}
\noindent\textbf{Transversely polarized scattering}

When both $e^-$ and $e^+$ beams are 100\% polarized along the $x$ direction, the spin state of the initial beams is a pure state with the state vector $\ket{e^-e^+}$ given in Eq.~\eqref{eq:statevector_intro}.  Multiplying the state vector with the transition matrix in Eq.~\eqref{eq:transitionmatrixMasslessTau}, we see that the final state is now a coherent superposition of the two states in Eq.~\eqref{eq:statevectorLR}, given by
\begin{align}\label{eq:tautauPurestate}
    \ket{f\bar f}&\propto \ket{f\bar f}_L + \ket{f\bar f}_R = \left(\mathcal{M}^{RL}_{\uparrow\uparrow}+\mathcal{M}_{\uparrow\uparrow}^{LR}\right)\ket{\uparrow\uparrow}  +\left(\mathcal{M}^{RL}_{\downarrow\downarrow}+\mathcal{M}_{\downarrow\downarrow}^{LR}\right)\ket{\downarrow\downarrow}.
\end{align}
The spin configuration of the final state is then determined by the ratio
\begin{equation}
\label{eq:X}
     X\equiv\frac{\mathcal{M}^{RL}_{\uparrow\uparrow}+\mathcal{M}_{\uparrow\uparrow}^{LR}}{\mathcal{M}^{RL}_{\downarrow\downarrow}+\mathcal{M}_{\downarrow\downarrow}^{LR}},
\end{equation}
where the explicit forms of the amplitudes are given in terms of chiral couplings in Eq.~\eqref{eq:transitionmatrixMasslessTau}, and we choose $\phi=0$ in the following for illustration.
A maximally entangled Bell state is produced if $|X|=1$, while a separable state is produced when either $X=0$ or $X=\infty$.

\subsubsection{ \texorpdfstring{$e^-e^+ \to \tau^- \tau^+ $}{ e- e+  ->  τ- τ+} }
\label{subsec:tautau}

Based on the above discussions and starting from unpolarized scattering, the entanglement of $\tau^-\tau^+$ from unpolarized scattering with different collision energies and scattering angles is shown in Fig.~\ref{fig:tauConcurrences}(a).  It is clearly seen that the phase space is divided into three regions by two horizontal lines $\sqrt{s}\approx 78$ GeV and $\sqrt{s}\approx 114$ GeV, in which the concurrence is zero; see also Ref.~\cite{Fabbrichesi:2024wcd,Guo:2026yhz}.
This is because Eq.~\eqref{eq:VeqA} is satisfied for the $e^-e^+\to \tau^-\tau^+$ process when 
\begin{equation}
\sqrt{s}= 2\sqrt{\frac{2s_W^2}{1+4s_W^2}}m_W \approx 78 {\rm ~GeV} \ \  \text{or}\ \  \sqrt{s}=\sqrt{2}m_W\approx 114{\rm ~GeV},\  (m_W= m_Z c_W),
\end{equation}
as also shown in the solid curves of Fig.~\ref{fig:fVfAtautau}(a).

In the low energy region, when $\sqrt{s}\sim 2m_\tau \ll m_Z$, the process is dominated by photon mediator with mostly vector current interactions. Similarly to the QED process, a maximally entangled triplet $\ket{\Psi}_{\hat n}$ is produced near the central scattering region. As energy increases and approaches $78~{\rm GeV}$, the contribution of axial vector current becomes comparable to the contribution of vector current, and entanglement as a function of the scattering angle develops a significant forward-backward asymmetry.
In the second energy region with $78~{\rm GeV}<\sqrt{s}<114~ {\rm GeV}$, the process is dominated by the axial current interaction of the $\tau$ leptons, and the final state is close to the spin triplet $\ket{\Psi}_{\hat r}$.
In the third energy region with $\sqrt{s}>\sqrt{2}m_W$, the vector current interaction of the $\tau$ lepton is back to dominant and the spin configuration is close to $\ket{\Psi}_{\hat n}$ again, in which case the axial vector current contribution introduces a forward-backward asymmetry but does not qualitatively change the spin configuration, similar to that of $e^-e^+\to t\bar{t}$.

\begin{figure}[tb]
	\centering
    \includegraphics[width=\linewidth]{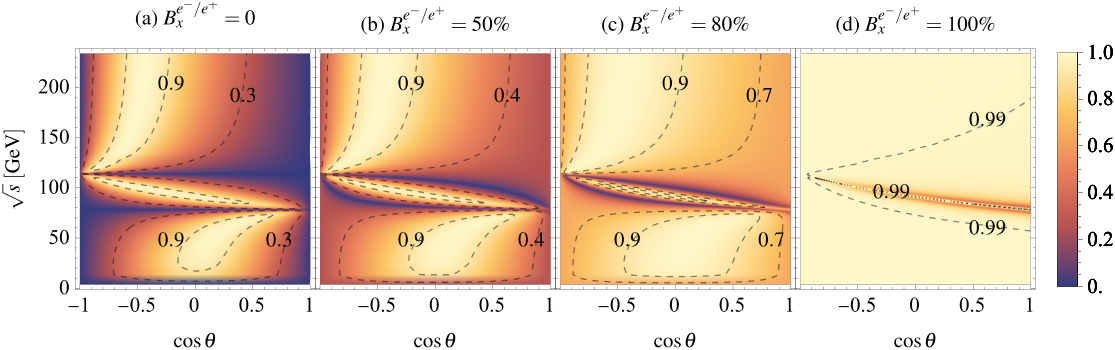}
	\caption{Concurrence of the spin density matrix for the EW process $e^-e^+\to \tau^-\tau^+$ in the $\cos\theta$-$\sqrt s$ plane in the $\tau^-\tau^+$ c.m.~frame with $\phi=0$: (a) unpolarized, (b) 50\% transversely polarized, (c) 80\% transversely polarized, and (d) 100\% transversely polarized beams.
    }
	\label{fig:tauConcurrences}
\end{figure}

When both initial beams are purely transversely polarized, we find that the  $\tau^-\tau^+$ pair is still very close to maximally entangled in most phase space regions even when the axial coupling is non-negligible, as shown in 
Fig.~\ref{fig:tauConcurrences}(d).
This is because numerically we have $(\mathcal{M}^{RL}_{\uparrow\uparrow}+\mathcal{M}_{\uparrow\uparrow}^{LR})\approx (\mathcal{M}^{RL}_{\downarrow\downarrow}+\mathcal{M}_{\downarrow\downarrow}^{LR})$.
From the explicit expression of amplitudes in Eq.~\eqref{eq:transitionmatrixMasslessTau}, the condition $(\mathcal{M}^{RL}_{\uparrow\uparrow}+\mathcal{M}_{\uparrow\uparrow}^{LR})= (\mathcal{M}^{RL}_{\downarrow\downarrow}+\mathcal{M}_{\downarrow\downarrow}^{LR})$ is satisfied when
\begin{align}\label{eq:ampdifference}
    \frac{e^2 s \left(1-4 s_W^2\right) }{4 c_W^2 s_W^2 \left(s-m_Z^2\right)} (1+\cos{\theta} )=0.
\end{align}
As long as the width effect is not significant, Eq.~\eqref{eq:ampdifference} is approximately satisfied due to the accidental cancellation in $(1/4-s_W^2)$.\footnote{In our calculation, we take $s_W^2=0.22225$.}
Consequently, when the scattering energy is away from the $Z$ threshold, either above or below, the spin state of the $\tau^-\tau^+$ pair at $\phi=0$ is approximately the spin triplet
\begin{equation}\label{eq:tautauPhin}
\ket{\Psi}_{\hat n}=\frac{\ket{\uparrow\uparrow}+\ket{\downarrow\downarrow}}{\sqrt{2}}.
\end{equation}
This $(1/4-s_W^2)$ suppression arises from the SM gauge couplings of charged leptons like $\tau$, and the numerical coincidence $X\approx 1$ generally does not hold for either top quark or bottom quark.  As we can see from the comparison among Figs.~\ref{fig:xxConcurrence}(d), ~\ref{fig:tauConcurrences}(d) and~\ref{fig:bbConcurrences}(d), with 100\% polarized initial beams, either $t\bar t$ pair or $b\bar b$ pair is produced less entangled than the $\tau^-\tau^+$ pair.

When the scattering energy is close to the $Z$ boson threshold, Eq.~\eqref{eq:ampdifference} is no longer valid because the resonance behavior and the $Z$-boson width effects need to be taken into account.
However, as calculated in Appendix~\ref{appendix:EW}, an exact solution of $|X|=1$ still exists, given by
\begin{equation}\label{eq:tautauX=1}
\sqrt{s}=\frac{4s_W m_W}{\sqrt{8s_W^2+1+\cos\theta}}, 
\end{equation}
which corresponds to the narrow band across $m_Z$ in Fig.~\ref{fig:tauConcurrences}(d).
As calculated in Appendix~\eqref{eqapp:tautauX=1}, in the limit of $\Gamma_Z\ll m_Z$, $X=-1$ so that the spin state is
\begin{equation}
     \ket{\Psi}_{\hat r}=\frac{\ket{\uparrow\uparrow}-\ket{\downarrow\downarrow}}{\sqrt{2}},
\end{equation}
a different triplet configuration from  Eq.~\eqref{eq:tautauPhin} that is produced in most other phase space regions.  We also see from the narrow band that the entanglement decreases when the spin state transitions between these two spin configurations; see Appendix~\ref{appendix:EW} for a more detailed discussion of the transition.

Figure~\ref{fig:tauConcurrences} also shows that the entanglement of the final states critically depends on the degree of transverse polarization of the initial beams. As given in Eq.~(\ref{eq:generalPolarization}), the final state is a mixture of the unpolarized case and the 100\% polarized case. As the degree of transverse polarization increases, the top and bottom entangled regions (with spin state close to $\ket{\Psi}_{\hat n}$) grow larger, whereas the middle entangled region (with spin state close to $\ket{\Psi}_{\hat r}$) shrinks.

\subsubsection{ \texorpdfstring{$e^-e^+ \to b \bar{b}$}{e- e+  ->  b bbar}}
\label{subsec:bb}

The general discussion in Eq.~\eqref{eq:VeqA} and Eq.~\eqref{eq:X} is valid for both $\tau$ leptons and $b$ quarks.
Therefore, the $e^-e^+\to b\bar{b}$ process has a qualitatively similar behavior to the $e^-e^+\to \tau^-\tau^+$ process, with some numerical differences due to different electroweak charges.

For unpolarized scattering, the entanglement of $b\bar{b}$ with different collision energies and scattering angles is shown in Fig.~\ref{fig:bbConcurrences}(a).  We see that the concurrence is zero when $\sqrt{s}\approx 67$ GeV and $\sqrt{s}\approx 165$ GeV, and the phase space is divided into three energy regions by these two horizontal lines, similar to the case of $\tau^-\tau^+$.
Again, this is because Eq.~\eqref{eq:VeqA} is satisfied when
\begin{align}
    &\sqrt{s}=m_W\sqrt{\frac{-3+\sqrt{9-48s_W^2+128s_W^4}}{-3+8s_W^2}}\approx67~{\rm GeV}\\
    \text{or}~~&\sqrt{s}=m_W\sqrt{\frac{3+\sqrt{9-48s_W^2+128s_W^4}}{-3+8s_W^2}}\approx 165~{\rm GeV},
\end{align}
as shown by the black line of Fig.~\ref{fig:fVfAtautau}(b).

Figure~\ref{fig:bbConcurrences}(a) also shows that the entanglement of $b\bar b$ from unpolarized scattering is much smaller than that of $\tau^-\tau^+$, which is due to significantly different axial/vector current interactions; see Fig.~\ref{fig:fVfAtautau}. Figure~\ref{fig:fVfAtautau}(a) shows that, for the $e^-e^+\to \tau^-\tau^+$ process, in most energy regions, either vector or axial current dominates the production of both $\ket{\tau^-\tau^+}_L$ and $\ket{\tau^-\tau^+}_R$. As a result, $\ket{\tau^-\tau^+}_L$ and $\ket{\tau^-\tau^+}_R$ are usually in the same triplet configuration and therefore their mixture exhibits greater entanglement. In contrast, for the $e^-e^+\to b\bar{b}$ process, Fig.~\ref{fig:fVfAtautau}(b) shows that usually one of $\ket{b\bar  b}_L$ and $\ket{b\bar b}_R$ is produced predominantly through the vector current interaction, while the other is dominated by the axial current.
Consequently, Eq.~\eqref{eq:LRmixing} is usually a mixture of two different configurations: one is closer to the triplet $\ket{\Psi}_{\hat n}$ while the other is closer to the triplet $\ket{\Psi}_{\hat r}$.  These two states have opposite spin correlation along both $\hat{n}$ and $\hat{r}$ directions, and their spin correlations cancel each other out in the mixture.  As we can see in Fig.~\ref{fig:bbConcurrences}(a), for unpolarized $e^-e^+$ scattering, the concurrence of $b\bar{b}$ in most phase space regions is much smaller than that of $\tau^-\tau^+$.

In the three energy regions in Fig.~\ref{fig:bbConcurrences}(a), the concurrence in the low-energy region ($\sqrt{s} < 67$ GeV) is relatively higher while the concurrence in the other two regions is significantly lower.
In the first region, when $\sqrt{s} \ll 67$ GeV, both $\ket{b\bar{b}}_L$ and $\ket{b\bar{b}}_R$ are dominated by vector current interactions. Similarly to the QED process, the entanglement increases with energy as the final state becomes relativistic, and both $\ket{b\bar{b}}_L$ and $\ket{b\bar{b}}_R$ become close to the spin state $\ket{\Psi}_{\hat n}$. However, when $\sqrt{s}\gtrsim 50$ GeV, $\ket{b\bar{b}}_L$ becomes close to the spin state $\ket{\Psi}_{\hat r}$ since it is dominated by the axial interaction, then the mixture between $\ket{b\bar b}_L$ and $\ket{b\bar b}_R$ destructs the entanglement and therefore the entanglement begins to decrease with energy.
In the second energy region, where $67 \text{ GeV} <\sqrt{s} < 165 \text{ GeV}$, the overall axial current dominates. However, in general, one of $\ket{b\bar{b}}_L$  and $\ket{b\bar{b}}_R$ is produced dominantly through the interaction of the vector current, while the other is dominated by the axial current.
Similarly, in the third energy region with $\sqrt{s} \gg 165$ GeV, although the vector current dominates on the whole, $\ket{b\bar{b}}_L$ is produced primarily through the vector current interaction while $\ket{b\bar{b}}_R$ is produced primarily through the axial current interaction. Thus, in most areas of these two high-energy regions, $\ket{b\bar{b}}_L$ and $\ket{b\bar{b}}_R$ are not in the same spin configuration, that is, one approaches $\ket{\Psi}_{\hat n}$ while the other approaches $\ket{\Psi}_{\hat r}$, and their mixture results in much smaller concurrence compared to the $e^-e^+\to\tau^-\tau^+$ process.

\begin{figure}
	\centering
\includegraphics[width=\linewidth]{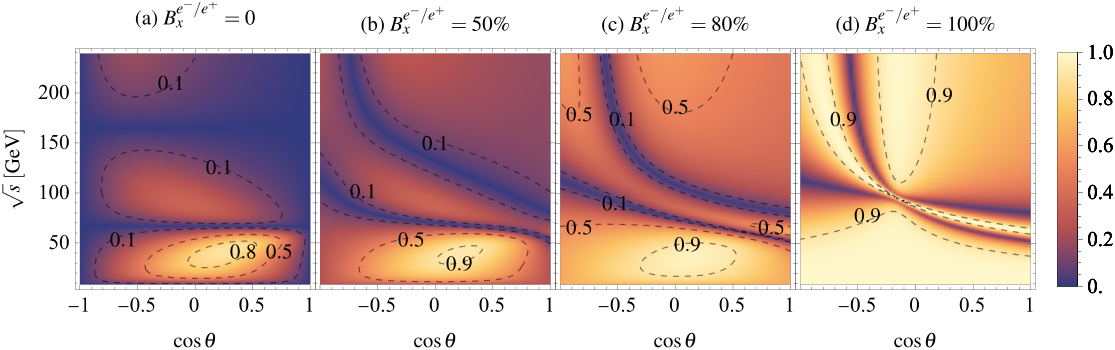}

	\caption{Concurrence of the spin density matrix for the EW process $e^-e^+\to b\bar{b}$ in the $\cos\theta$-$\sqrt s$ plane in the $b\bar{b}$ c.m.~frame with $\phi=0$: (a) unpolarized, (b) 50\% transversely polarized, (c) 80\% transversely polarized, and (d) 100\% transversely polarized beams.
    }
	\label{fig:bbConcurrences}
\end{figure}

When both $e^-$ and $e^+$ beams are 100\% polarized along $x$ direction, the final $b\bar b$ state is a pure state given by Eq.~\eqref{eq:tautauPurestate},
and the spin configuration of $b\bar{b}$ is determined by $X=(\mathcal{M}^{RL}_{\uparrow\uparrow}+\mathcal{M}_{\uparrow\uparrow}^{LR})/(\mathcal{M}^{RL}_{\downarrow\downarrow}+\mathcal{M}_{\downarrow\downarrow}^{LR})$.
Due to different chiral couplings, for the $b\bar b$ final state, the condition $\mathcal{M}^{RL}_{\uparrow\uparrow}+\mathcal{M}_{\uparrow\uparrow}^{LR}\approx \mathcal{M}^{RL}_{\downarrow\downarrow}+\mathcal{M}_{\downarrow\downarrow}^{LR}$ is not always satisfied as in the case of the $\tau$ lepton.
However, maximal entanglement can still happen in some phase space region.
There exists an exact solution of $|X|=1$ even after considering the width effect, given by
\begin{equation} \label{eq:bbX=1}
  \sqrt{s}=\frac{4m_Ws_W}{\sqrt{3+3\cos{\theta}}},
\end{equation}
which corresponds to the maximum entangled yellow band in Fig.~\ref{fig:bbConcurrences}(d).

In other phase space regions than Eq.~\eqref{eq:tautauX=1}, the $b\bar b$ is usually not as entangled as $\tau^-\tau^+$.
Moreover, in some phase space region the $b\bar b$ could be separable if
\begin{equation}
    \mathcal{M}^{RL}_{\uparrow\uparrow}+\mathcal{M}_{\uparrow\uparrow}^{LR}=0 \text{~~~~or~~~~} \mathcal{M}^{RL}_{\downarrow\downarrow}+\mathcal{M}_{\downarrow\downarrow}^{LR}=0.
\end{equation}
When the effect of $Z$-width is negligible, this happens if
\begin{equation}\label{eq:XX_bbseparable}
    \sqrt{s}=\frac{2m_W}{\sqrt{3+\cos{\theta}}}\quad \mathrm{or}\quad\sqrt{s}=\frac{2\sqrt{2}m_Ws_W}{\sqrt{3-6s_W^2+(3-2s_W^2)\cos{\theta}}},
\end{equation}
which corresponds to the two narrow blue bands in Fig.~\ref{fig:bbConcurrences}(d).
In these two narrow blue bands the final state $b\bar b$ is separable, while in most other phase space regions the $b\bar b$ is still largely entangled.

\subsection{Quantum magic}

The quantum magic of the final state in electro-weak processes can be analyzed similarly to the spin state summarized above. For the $t\bar t$ process, the overall characteristic is still very similar to that of the QED process.
As shown in Fig.~\ref{fig:magicTT}, with both transversely polarized beams, the quantum magic is very small when $\phi=0$.
A sizable magic occurs only for events with a different azimuthal angle, where the spin state is still largely entangled but diagonal in a different basis than the helicity basis. 

\begin{figure}
    \centering
    \includegraphics[scale=1]{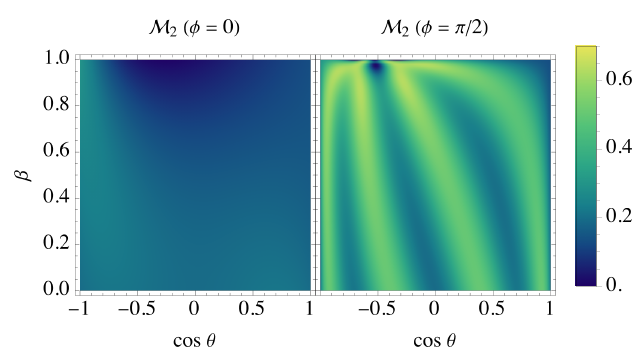}
    \caption{Quantum magic (SSRE) for the EW process $e^-e^+ \to  t \bar t$ in the $\cos\theta$-$\beta$ plane for the  100\% transversely polarized beams at (a) $\phi=0$ and (b) $\phi=\pi/2$.}
    \label{fig:magicTT}
\end{figure}

For $\tau^-\tau^+$ pair and $b\bar b$ pair production, the ratio between vector and axial vector current coupling changes dramatically around the $Z$ pole, leading to a variation of the spin configuration as discussed in previous sections. Therefore, unlike the QED process, magic can still be non-zero at $\phi=0$ as shown in Fig.~\ref{fig:magictautauBB}. Compared with Fig.~\ref{fig:tauConcurrences} and Fig.~\ref{fig:bbConcurrences}, we see that the magic minimizes in the maximal entangled region given by Eq.~\eqref{eq:tautauX=1} and Eq.~\eqref{eq:bbX=1}, where the state is approximately the stabilizer state $\ket{\uparrow\uparrow}\pm\ket{\downarrow\downarrow}$. For $b\bar b$, the magic is also zero when Eq.~\eqref{eq:XX_bbseparable} is satisfied, in which case the state is a completely separable state $\ket{\uparrow\uparrow}$ or $\ket{\downarrow\downarrow}$.

\begin{figure}
    \centering
    \includegraphics[scale=1]{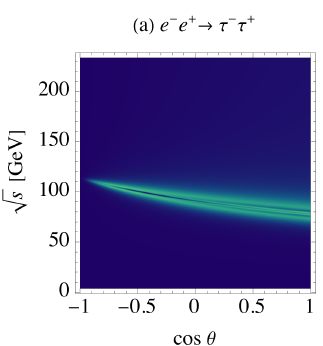}~~~~~~
    \includegraphics[scale=1]{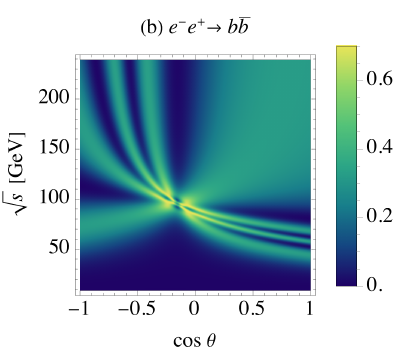}
    \caption{Quantum magic (SSRE) for the EW processes (a) $e^-e^+ \to \tau^-\tau^+$ and (b) $e^-e^+ \to b\bar b$ in the $\cos\theta$-$\sqrt s$ plane for the 100\% transversely polarized beam at $\phi=0$.}
    \label{fig:magictautauBB}
\end{figure}

\section{Discussions and Conclusion}
\label{section:summary}

In this work, we have systematically investigated the influence of transversely polarized initial beams on the spin correlations and quantum entanglement of fermion pairs ($f\bar{f}$) produced in $e^-e^+$ annihilation at the leading order. Since interactions govern the formation of the quantum states, we consider both QED processes and electroweak processes in the kinematic approach \cite{Cheng:2024rxi} with various representative final states including $t\bar{t}$, $\tau^-\tau^+$, and $b\bar{b}$.

Our key finding is that transverse polarization of the electron and positron beams has a dramatic effect on the quantum entanglement of the final state.
In particular, for a pure QED process, we found that the fermion pair produced from the annihilation of $100\%$ transversely polarized $e^\pm$ beams is in a maximally entangled Bell state across the entire phase space, independent of the center-of-mass energy and the scattering angle, as shown in Fig.~\ref{fig:QEDConcurrence}(d). This is in stark contrast to the unpolarized or longitudinal polarized case, where significant entanglement is only observed in the ultra-relativistic, central scattering region.
This remarkable feature was elucidated both through explicit calculation of the spin density matrix in Eq.~\eqref{eq:Cijxx} and by identifying a ``diagonal basis" in which the scattering amplitudes are simplified, revealing the maximal entanglement at the amplitude level in Section~\ref{sub:diagonalBasis}.
We also observed that quantum magic measures such as SSRE could be non-zero in the helicity basis even if the state is maximally entangled, as shown in Fig.~\ref{fig:QEDmagic}, which is a result of the angular dependence of the spin state $\ket{\Psi}_{\hat e_2}$ in Eq.~\eqref{eq:QEDdiagonalbasis}.

For electroweak processes with both unpolarized and transversely polarized initial beams, we investigated how the entanglement of the final states depends on chiral couplings in detail.  We showed that the ratio between the axial and vector current coupling strength is the major factor in determining the entanglement of the final states.
For $e^-e^+ \to t\bar{t}$, the axial-vector interaction in the exchange of $Z$ bosons slightly reduces the concurrence, but the entanglement behavior remains qualitatively similar to the QED case because the process is dominated by the vector current interaction, as shown in Figs.~\ref{fig:fVfA} and~\ref{fig:xxConcurrence}.
The processes $e^-e^+\to \tau^-\tau^+$ and $e^-e^+\to b\bar b$ can occur below or above the $Z$ pole, and the entanglement between final states exhibits a much richer structure because the ratio between axial and vector current coupling strength significantly depends on the scattering energy near the $Z$ pole, as shown in Fig.~\ref{fig:fVfAtautau}. In the unpolarized case, entanglement vanishes at specific energies when the vector and axial vector couplings of the final state contribute equally to the process, i.e., $(f^L_V)^2+ (f^R_V)^2=
    (f^L_A)^2+ (f^R_A)^2$.
With fully transversely polarized beams, the spin configuration and entanglement of the final state sensitively depend on the values of chiral couplings.  The $\tau$ lepton pair is produced in a nearly maximally entangled state in most phase space regions as shown in Fig.~\ref{fig:tauConcurrences}(d), due to the accidental cancellation of the charged-lepton neutral current coupling $1-4s_W^2\approx 0$. With a different gauge quantum number and chiral coupling, $b\bar{b}$ is usually produced in a less entangled spin state at various phase space points in Fig.~\ref{fig:bbConcurrences}(d). For a comparative study, we consider quantum magic in EW processes. Consistent with our analysis, the SSRE of the $t\bar t$ process is found to be similar to that of QED processes as shown in Fig.~\ref{fig:magicTT}, while the SSRE of $\tau^-\tau^+/b\bar b$ is more featureful and the magic maximizes in the phase space regions where the state transitions between different stabilizer spin configurations due to varying chiral couplings, as shown in Fig.~\ref{fig:magictautauBB}.

Two final remarks are in order. 
Firstly, unlike the $s$-channel processes discussed so far,  the $t$-channel contributions to the fermion pair production, such as in Bhabha scattering $e^-e^+\to e^-e^+$, can dilute the spin correlations and the maximal entanglement is not always preserved. 
Since the initial beams with arbitrary polarizations can interact with each other via the $t$-channel exchange, the role of transverse polarization is no longer simply to provide the coherent superposition $\ket{RL}+\ket{LR}$ in Eq.~\eqref{eq:statevector_intro} to participate in the interaction. Instead, the transverse polarization changes the way different diagrams interfere with each other. This has been shown even for longitudinal polarizations that can significantly change the entanglement between the final-state particles~\cite{Guo:2026yhz}. Nevertheless, the final spin state still sensitively depends on the transverse polarization, as shown in Fig.~\ref{fig:eeConcurrences}, qualitatively following that of Fig.~\ref{fig:tauConcurrences}. 

\begin{figure}
	\centering
\includegraphics[width=\linewidth]{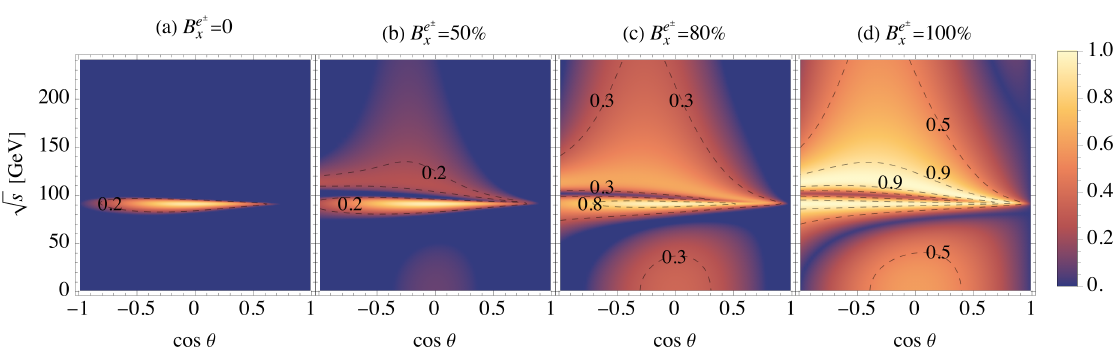}

	\caption{Concurrence of final state for the EW process $e^-e^+\to e^-e^+$ in the $\cos\theta$-$\sqrt s$ plane in the c.m.~frame with $\phi=0$: (a) unpolarized, (b) 50\% transversely polarized, (c) 80\% transversely polarized, and (d) 100\% transversely polarized beams.
    }
	\label{fig:eeConcurrences}
\end{figure}

Secondly, when considering angular-averaged states in a finite phase space region $\Pi$,
\begin{equation}
    \bar\rho_{\Pi}=\frac{1}{\sigma_\Pi}\int_{\Omega\in\Pi}d\Omega\frac{d\sigma}{d\Omega}\rho(\mathbf{k})_{\alpha\bar\alpha,\alpha'\bar\alpha'}\ket{\alpha\bar\alpha}\bra{\alpha'\bar\alpha'}, \quad \sigma_\Pi\equiv\int_{\Pi}d\Omega \frac{d\sigma}{d\Omega},
\end{equation}
the transverse polarization of the initial state leads to additional basis-dependence on the azimuthal angle.
The basis-dependence of fictitious states has been extensively explored for unpolarized or longitudinally polarized beams~\cite{Cheng:2023qmz,Cheng:2024btk,CMS:2025brx}.
In these cases, the average over azimuthal angle $\phi$ is rather trivial in the helicity basis, because the scattering process and the final spin state exhibit an azimuthal rotation symmetry. 
However, the transversely polarized initial beams break the azimuthal rotation symmetry, and the azimuthal angular average is a nontrivial step to perform.
As an illustration, in Fig.~\ref{fig:QEDaverage}, we show the concurrence of the azimuthal angular-average of the state $\ket{\Psi}_{\hat e_2}$ in Eq.~\eqref{eq:QED_BellStateVector} produced in the QED process.
Unlike unpolarized scattering, the average over the azimuthal angle in both the helicity basis and the fixed beam basis leads to cancellation of spin correlation to a certain extent.
This is because the spin state $\ket{\Psi}_{\hat e_2}$ depends on $\phi$ and the mixture between different Bell states in Eq.~\eqref{eq:Bell_state} cancels the spin correlation. 
The diagonal basis, on the other hand, is optimal to maintain the full spin-correlation produced from transversely polarized beams.}

\begin{figure}
    \centering
    \includegraphics[width=0.5\linewidth]{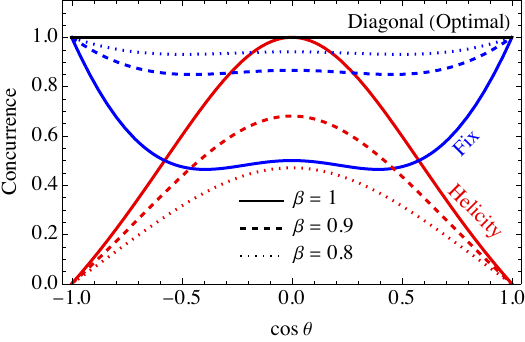}
    \caption{Concurrence of the azimuthal angular averaged of the spin triplet $\ket{\Psi}_{\hat e_2}$ (Eq.~\ref{eq:QED_BellStateVector}) versus $\cos\theta$ in fixed beam basis (blue curves), helicity basis (red curves), and diagonal basis (black straight line), with a variety of speed values.}
    \label{fig:QEDaverage}
\end{figure}

Overall, by investigating how initial beam polarization affects the quantum state produced from electron-positron annihilation, we showed that the transverse beam polarization is a powerful tool for generating and controlling quantum entanglement in collider experiments. It enables the production of maximally entangled states throughout the phase space in QED processes and significantly enhances the entanglement in electroweak processes. These findings open new opportunities for quantum state engineering at high-energy colliders and suggest that future experiments with polarized beams, such as those at the Super Charm-Tau factory~\cite{Bao:2025aic}, FCC-ee~\cite{FCC:2018evy}, and CEPC~\cite{CEPCStudyGroup:2023quu}, could serve as powerful laboratories for quantum information science.

\acknowledgments
This work was supported in part by the US Department of Energy under grant No.~DE-SC0007914 and in part by Pitt PACC.
The work of YJF and HZ is supported in part by the National Science Foundation of China under 
Grants No.~12575114, No.~12235001.
The work of YDL is supported in part by the National Science Foundation of China under Grant Nos. 12075257, 12175016, the National Key R$\&$D Program of China under Grant No. 2023YFA1607104 and Fundamental Research Funds for the Central Universities, Beijing
Normal University.

\appendix
\section{Partially polarized initial states}
In this appendix, we show that when the initial state is partially transversely polarized ($|B_x^{e^\pm}|$), the final state density matrix is a mixture between the unpolarized case and purely transversely polarized case, and we present the explicit form of the density matrix as a function of initial state polarization.

\subsection{Effective initial states that participate in the interaction}

In the massless limit, only electrons and positrons with opposite helicities (parallel spins) annihilate with each other, then it is convenient to write down the effective $e^-e^+$ state that participates in the interaction.

The density matrix of the initial beam can be expressed in the basis  $\{\ket{RL},\ket{RR}, \ket{LL},\allowbreak \ket{LR}\}$ and only the submatrix in the subspace spanned by  $\{\ket{RL},\ket{LR}\}$ contributes to the process.
For example, the density operator of an unpolarized $e^-e^+$ pair is
\begin{align}
    \hat{\rho}_{\rm un}^{e^-}\otimes
    \hat{\rho}_{\rm un}^{e^+} &= \left(\frac{\ket{R}\bra{R}+\ket{L}\bra{L}}{2}\right)\otimes \left(\frac{\ket{R}\bra{R}+\ket{L}\bra{L}}{2}\right) \nonumber\\
    &=\frac{\ket{RL}\bra{RL}+\ket{RR}\bra{RR}+ \ket{LL}\bra{LL} + \ket{LR}\bra{LR}}{4}, \\
 \xrightarrow{\text{interacting  part}} \quad \hat\rho^{e^-e^+}_{\rm un, eff} &=\frac{\ket{RL}\bra{RL} + \ket{LR}\bra{LR}}{4},
\end{align}
where $\hat\rho^{e^-e^+}_{\rm un, eff} $ is the effective part of $\hat{\rho}_{\rm un}^{e^-}\otimes \hat{\rho}_{\rm un}^{e^+}$ that participates in the interaction.
Another example is the case that both $e^-$ and $e^+$ are 100\% transversely polarized along $x$ direction, the density matrix is 
\begin{align}
    \hat{\rho}_{xx}^{e^-}\otimes
    \hat{\rho}_{xx}^{e^+} &= \left[\left(\frac{\ket{R}+\ket{L}}{\sqrt{2}} \right)\left(\frac{\bra{R}+\bra{L}}{\sqrt{2}}\right)\right]\otimes \left[\left(\frac{\ket{R}+\ket{L}}{\sqrt{2}} \right)\left(\frac{\bra{R}+\bra{L}}{\sqrt{2}}\right)\right]
\end{align}
and the effective part that participates in the interaction is
\begin{equation}
\hat{\rho}_{xx}^{e^-}\otimes
    \hat{\rho}_{xx}^{e^+} \xrightarrow{\text{interacting part}} \hat\rho^{e^-e^+}_{xx,\text{eff}} =\frac{1}{4}\left(\ket{RL} + \ket{LR}\right)\left(\bra{RL} + \bra{LR}\right).
\end{equation} 

Since the other parts in the density matrix are irrelevant to the interaction, the final state is only determined by the effective density operator:
\begin{equation}
    \hat{R}^{f\bar f}=\hat{\mathcal{T}} (\hat{\rho}^{e^-}\otimes \hat{\rho}^{e^+})\hat{\mathcal{T}}^\dagger=
    \hat{\mathcal{T}}~ \hat{\rho}^{e^-e^+}_{\rm eff}~\hat{\mathcal{T}}^\dagger
\end{equation}
where $\hat{R}_{f\bar f}$ is the unnormalized density operator of the final states $f\bar f$.

For partially transversely polarized initial beams as in Eq.~\eqref{eq:rhoxab}, we see that the effective part of the density matrix that participates in the interaction is given by
\begin{align}\label{eq:rhoeffBxBx}
    \hat{\rho}_{\rm eff}^{e^-e^+} &= \frac{1}{4}\big(\ket{RL}\bra{RL}+\ket{LR}\bra{LR}\big)+\frac{1}{4}B_x^{e^-}B_x^{e^+}\big(\ket{RL}\bra{LR}+\ket{LR}\bra{RL}\big) \nonumber\\
    &= (1-B^{e^-}_x B^{e^+}_x) 
    \hat{\rho}_{\rm un, eff}^{e^-e^+} +B^{e^-}_x B^{e^+}_x \hat{\rho}_{xx,\rm eff}^{e^-e^+}, 
\end{align}
which is a mixture of the effective density operator of unpolarized beams and 100\% transversely polarized beams.
Then, the unnormalized density operator of the final state is given by
\begin{align}\label{eqapp:generalPolarization}
    \hat{R}^{f\bar f}
    &= \hat{\mathcal{T}}~ \hat{\rho}^{e^-e^+}_{\rm eff}~\hat{\mathcal{T}}^\dagger \nonumber\\
    &= (1-B^{e^-}_x B^{e^+}_x) \hat R^{(\rm un)} + B^{e^-}_x B^{e^+}_x \hat R^{(xx)}
\end{align}
where $ \hat{R}^{(\rm un)} = \hat{\mathcal{T}}~ \hat{\rho}^{e^-e^+}_{\rm un, eff}~\hat{\mathcal{T}}^\dagger $ is the unnormalized spin density operator of the final states produced from the unpolarized initial state, and $ \hat{R}^{(xx)} = \hat{\mathcal{T}}~ \hat{\rho}^{e^-e^+}_{{xx,\rm  eff}}~\hat{\mathcal{T}}^\dagger $ is the unnormalized spin density matrix of the final state produced from the 100\% transversely polarized initial states. This applies to both the QED process and the EW process.
Normalizing Eq.~\eqref{eqapp:generalPolarization} obtains Eq.~\eqref{eq:generalPolarization}.

It is interesting to note that as long as one beam in the initial state is unpolarized, the effective density operator of $e^-e^+$ is reduced to the unpolarized case, and the final state is not different from unpolarized scattering even if another beam is transversely polarized.

\subsection{Explicit expression for final state density matrix}
\label{appendix:genneral_Cij}

Here we list the explicit expressions of the final state density matrices for the QED processes.
The unnormalized density matrix $\hat{R}^{(\rm un)}$ for unpolarized scattering and $\hat{R}^{(xx)}$ for 100\% transversely polarized scattering are
\begin{equation}
\begin{aligned}
    R^{(\rm un)}_{\alpha\bar\alpha,\alpha'\bar\alpha'}&= \mathcal{M}_{\alpha\bar\alpha}^{RL}(\mathcal{M}^{RL}_{\alpha'\bar\alpha'})^{\dagger}+\mathcal{M}_{\alpha\bar\alpha}^{LR}(\mathcal{M}^{LR}_{\alpha'\bar\alpha'})^{\dagger}\\
    &=e^4\begin{pmatrix}
        \cos{(2\theta)}+3 &-2c_\theta s_\theta \sqrt{1-\beta^2} &-2c_\theta s_\theta \sqrt{1-\beta^2} &2s_\theta^2\\
        -2c_\theta s_\theta \sqrt{1-\beta^2} &2(1-\beta^2)s_\theta^2 &2(1-\beta^2)s_\theta^2 &2c_\theta s_\theta \sqrt{1-\beta^2} \\
        -2c_\theta s_\theta \sqrt{1-\beta^2} &2(1-\beta^2)s_\theta^2 &2(1-\beta^2)s_\theta^2 &2c_\theta s_\theta \sqrt{1-\beta^2} \\
    2s_\theta^2&2c_\theta s_\theta \sqrt{1-\beta^2} &2c_\theta s_\theta \sqrt{1-\beta^2} &\cos{(2\theta)}+3 \\
    \end{pmatrix}.
    \label{eq:rho_un}
\end{aligned}
\end{equation}
and
\begin{equation}
\begin{aligned}
    R^{(xx)}_{\alpha\bar\alpha,\alpha'\bar\alpha'}&=\mathcal{M}_{\alpha\bar\alpha}^{RL}(\mathcal{M}_{\alpha'\bar\alpha'}^{RL})^{\dagger}+\mathcal{M}_{\alpha\bar\alpha}^{LR}(\mathcal{M}_{\alpha'\bar\alpha'}^{LR})^{\dagger}+ \mathcal{M}_{\alpha\bar\alpha}^{RL}(\mathcal{M}_{\alpha'\bar\alpha'}^{LR})^{\dagger}+\mathcal{M}_{\alpha\bar\alpha}^{LR}(\mathcal{M}_{\alpha'\bar\alpha'}^{RL})^{\dagger}\\
    &=4e^4\begin{pmatrix}
       c_\theta^2 s_\phi^2 +c_\phi^2 &-X_1 &-X_1 &(c_\phi + ic_\theta s_\phi)^2 \\
      -X_2 & (1-\beta^2)s_\theta^2 s_\phi^2 &(1-\beta^2)s_\theta^2 s_\phi^2 & X_1 \\
      -X_2 & (1-\beta^2)s_\theta^2 s_\phi^2 &(1-\beta^2)s_\theta^2 s_\phi^2 & X_1\\
       (c_\phi - ic_\theta s_\phi)^2 &X_2 &X_2 &c_\theta^2 s_\phi^2 +c_\phi^2\\
    \end{pmatrix},
    \label{eq:rho_xx}
\end{aligned}
\end{equation}
where $X_1=s_\phi s_\theta \sqrt{1-\beta^2}(c_\theta s_\phi - i c_\phi)$,$ X_2=s_\phi s_\theta \sqrt{1-\beta^2}(c_\theta s_\phi + i c_\phi)$.

Then the normalized density matrix for the  generally transversely polarized initial state is,
\begin{equation}
    \begin{aligned}
        \rho^{f \bar f}(\mathbf{k})_{\alpha\bar\alpha,\alpha'\bar\alpha'}&= \frac{R^{f\bar f}(\mathbf{k})_{\alpha\bar\alpha,\alpha'\bar\alpha'}}{\tr(R^{f\bar f}(\mathbf{k}))},\\
        &=\frac{N_1}{N}(1-B_x^{e^-}B_x^{e^+})\rho_{\alpha\bar\alpha,\alpha'\bar\alpha'}^{\rm (un)}+\frac{N_2}{N}B_x^{e^-}B_x^{e^+}\rho_{\alpha\bar\alpha,\alpha'\bar\alpha'}^{(xx)},\\
    \end{aligned}
\end{equation}
where 
\begin{align}
  N_1&=\tr(R^{(\rm un)})\propto d\sigma^{(\rm un)},\\
   N_2&=\tr(R^{(xx)})\propto d\sigma^{(xx)},\\
   N&=(1-B_x^{e^-}B_x^{e^+}) N_1 +B_x^{e^-}B_x^{e^+} N_2 \nonumber\\
   & \propto (1-B_x^{e^-}B_x^{e^+}) d\sigma^{(\rm un)} +B_x^{e^-}B_x^{e^+} d\sigma^{(xx)}=d\sigma^{ff}
\end{align}
and for the QED process, 
\begin{equation}
    \begin{aligned}
        &N_1=4 e^4\left(2-\beta^2 s_\theta^2 \right),\\
        &N_2= 4 e^4\left( 2- \beta^2 s_\theta^2   +\beta^2 s_\theta^2 \cos(2\phi)\right).
    \end{aligned}
\end{equation}
As discussed before, the final state density matrix for the generally transversely polarized initial state is a mixture of the final state produced from unpolarized initial states and the final state produced from 100\% transversely polarized initial states, and the weight in the mixture is given by both the degree of polarization and the differential cross section,
\begin{equation}
     \rho^{f\bar f}_{\alpha\bar\alpha,\alpha'\bar\alpha'}=\frac{d\sigma^{(\rm un)}}{d\sigma^{f\bar f}}(1-B_x^{e^-}B_x^{e^+})\rho_{\alpha\bar\alpha,\alpha'\bar\alpha'}^{(\rm un)}+\frac{d\sigma^{(xx)}}{d\sigma^{f\bar f}}B_x^{e^-}B_x^{e^+}\rho_{\alpha\bar\alpha,\alpha'\bar\alpha'}^{(xx)}.
\end{equation}

Since the spin correlation matrix is linearly related to the normalized density matrix, the spin correlation matrix of this general transversely polarized state can be written as a combination of $C_{ij}^{\rm (un)}$ and $C_{ij}^{(xx)}$,
\begin{equation}
    C_{ij}=\frac{N_1}{N}\left(1-B_x^{e^-}B_x^{e^+}\right)C_{ij}^{\rm(un)}+\frac{N_2}{N}B_x^{e^-}B_x^{e^+}C_{ij}^{(xx)}.
\end{equation}
The three eigenvalues of $C_{ij}$ are
\begin{subequations}
\begin{eqnarray}
{{c_1}}&{{=}}&{{1,}}\\
{{c_2}}&{{=}}&{{\sqrt{1-\frac{4[1-(B_x^{e^-}B_x^{e^+})^2](1-\beta^2\sin^2\theta)}{(2-\beta^2\sin^2\theta+B_x^{e^-}B_x^{e^+}\beta^2\sin^2\theta\cos2\phi)^2}},}}\\
{{c_3}}&{{=}}&{{-\sqrt{1-\frac{4[1-(B_x^{e^-}B_x^{e^+})^2](1-\beta^2\sin^2\theta)}{(2-\beta^2\sin^2\theta+B_x^{e^-}B_x^{e^+}\beta^2\sin^2\theta\cos2\phi)^2}},}}
\end{eqnarray}
\end{subequations}
so the concurrence is
\begin{equation}
    \mathscr{C}[\hat\rho]=\sqrt{1-\frac{4[1-(B_x^{e^-}B_x^{e^+})^2](1-\beta^2\sin^2\theta)}{(2-\beta^2\sin^2\theta+B_x^{e^-}B_x^{e^+}\beta^2\sin^2\theta\cos2\phi)^2}}.
\end{equation}

\section{Spin states of massless fermion pairs in EW processes}
\label{appendix:EW}

 The transition matrix in the massless limit is
\begin{align}
  \mathcal{T}_{\alpha\bar\alpha,\lambda\bar\lambda}= \begin{pmatrix}
        \mathcal{M}^{RL}_{\uparrow\uparrow} &0 &0 & \mathcal{M}^{LR}_{\uparrow\uparrow} \\
        0 &0 &0 & 0 \\
        0 &0 &0 & 0 \\
        \mathcal{M}^{RL}_{\downarrow\downarrow} &0 &0 & \mathcal{M}^{LR}_{\downarrow\downarrow} \\
    \end{pmatrix}&
    = \begin{pmatrix}
       (f^R_V+f^R_A) \frac{1+\cos\theta}{2}e^{i\phi} &0 &0 &  (f^L_V+f^L_A)\frac{1-\cos\theta}{2}e^{-i\phi} \\
        0&0&0&0\\
        0&0&0&0\\
        (f^R_V-f^R_A) \frac{1-\cos\theta}{2}e^{i\phi} &0 &0 &  (f^L_V-f^L_A) \frac{1+\cos\theta}{2}e^{-i\phi} \\
    \end{pmatrix}\\
    &= \begin{pmatrix}
       f^R_R \frac{1+\cos\theta}{2}e^{i\phi} &0 &0 &  f^L_R\frac{1-\cos\theta}{2}e^{-i\phi} \\
        0&0&0&0\\
        0&0&0&0\\
        f^R_L \frac{1-\cos\theta}{2}e^{i\phi} &0 &0 &  f^L_L \frac{1+\cos\theta}{2}e^{-i\phi} \\
    \end{pmatrix}.\nonumber
\end{align}
When the initial state is 100\% transversely polarized, the spin state vector is $\psi^{ee}_{\lambda\bar\lambda}\propto(1,1,1,1)$ and the spin state vector of $f\bar f$ at $\phi=0$ is
\begin{align}
    \ket{f\bar f} 
    &\propto \left(\mathcal{M}^{RL}_{\uparrow\uparrow}+\mathcal{M}_{\uparrow\uparrow}^{LR}\right)\ket{\uparrow\uparrow}  +\left(\mathcal{M}^{RL}_{\downarrow\downarrow}+\mathcal{M}_{\downarrow\downarrow}^{LR}\right)\ket{\downarrow\downarrow} \nonumber \\
    &\propto (f^V_R+f^A_R\cos\theta)\ket{\uparrow\uparrow}
    + (f^V_L-f^A_L\cos\theta)\ket{\downarrow\downarrow}.
\end{align}
where 
\begin{equation}
        f^V_L=\frac{f^R_L+f^L_L}{2},~~
        f^V_R=\frac{f^R_R+f^L_R}{2},~~
        f^A_L=\frac{f^R_L-f^L_L}{2},~~
        f^A_R=\frac{f^R_R-f^L_R}{2}.
\end{equation}

We see clearly that the ratio
\begin{equation}
    X=\frac{f^V_R+f^A_R\cos\theta}{f^V_L-f^A_L\cos\theta}
\end{equation}
determines the spin state. The absolute value $|X|$ determines the entanglement of the state while the complete phase of $X$ determines the spin configuration, e.g., $X=\pm1$ corresponds to different types of triplet.

\begin{figure}
    \centering
    \includegraphics[width=0.6\linewidth]{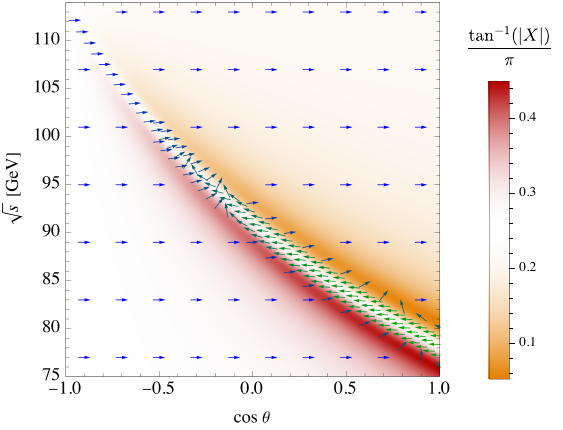}
    \includegraphics[width=0.6\linewidth]{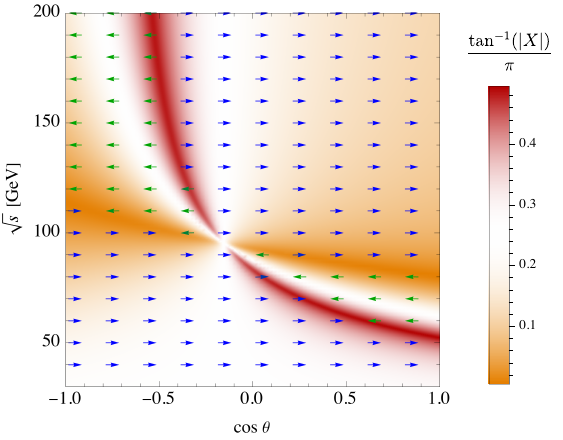}
    \caption{Value and phase of $X$ as a function of scattering angle and energy for $e^-e^+\to \tau^-\tau^+$ (upper) and $e^-e^+\to b\bar b$ processes (lower). The background color denotes the value of $|X|$ while the arrow direction denotes the complex phase of $X$.  In the white region ($|X|=1$), the state is maximally entangled and the forward/backward arrow corresponds to $X=\pm1$ giving the state triplet $\ket{\Psi}_{\hat n/\hat r}$.  In the dark red/orange region the state is close to the separable state $\ket{\uparrow\uparrow}/\ket{\downarrow\downarrow}$. }
    \label{fig:Xphases}
\end{figure}

If the ratio $|X|=1$, then the final state is a maximally entangled state.
For $\tau^-\tau^+$ production, an exact solution of $|X|=1$ is given by
\begin{equation}\label{eqapp:tautauX=1}
\sqrt{s}=\frac{4s_W m_W}{\sqrt{8s_W^2+1+\cos\theta}}, \quad X=-\frac{1+\cos\theta-i \frac{\Gamma_Z}{m_Z}\frac{16s_W^2 c_W^2}{1-4s_W^2}}{1+\cos\theta+i \frac{\Gamma_Z}{m_Z}\frac{16s_W^2 c_W^2}{1-4s_W^2}},
\end{equation}
We also see that the phase of $X$ is changing continuously at this line of $|X|=1$, with $X\approx -1$ when $\cos\theta=1$ and $X=1$ when $\cos\theta=-1$. In Fig.~\ref{fig:Xphases}(a) we show the absolute value and the complex phase of $X$, where he background color denotes the value of $|X|$ while the arrow direction denotes the complex phase of $X$.
We see the spin configuration changes while maintaining a Bell state in Eq.~\eqref{eqapp:tautauX=1}.

Similarly, for $b\bar b$ production, an exact solution of $|X|=1$ is given by
\begin{equation}
  \sqrt{s}=\frac{4m_Ws_W}{\sqrt{3+3\cos{\theta}}}, \quad X=-\frac{3(1-4s_W^2) +(3-4s_W^2)\cos\theta +i \frac{\Gamma_Z}{m_Z}\frac{16s_W^2 c_W^2}{1-4s_W^2}}{3(1-4s_W^2) +(3-4s_W^2)\cos\theta - i \frac{\Gamma_Z}{m_Z}\frac{16s_W^2 c_W^2}{1-4s_W^2}},
\end{equation}
and we show the absolute value and the complex phase of $X$ in Fig.~\ref{fig:Xphases}(b).

\clearpage
\bibliographystyle{JHEP}
\bibliography{ref}
\end{document}